\documentclass[fleqn,usenatbib]{mnras}

\usepackage{newtxtext,newtxmath}

\usepackage[T1]{fontenc}

\usepackage{graphicx}	
\usepackage{amsmath}	
\usepackage[normalem]{ulem}
\usepackage{booktabs}

\newcommand{\kms}{km/s~}
\newcommand{\Vpeak}{$V_{\rm{peak}}$~}
\newcommand{\cmg}{cm$^2$g$^{-1}$~}

\title[SIDM Subhalos for moderate cross sections]{
Motivations for a Large Self-Interacting Dark Matter Cross Section from Milky Way Satellites
}

\author[M. Silverman et al.]{
Maya Silverman,$^{1}$\thanks{E-mail: msilver2@uci.edu}
James S. Bullock,$^{1}$
Manoj Kaplinghat,$^{1}$
Victor H. Robles,$^{1,2}$
Mauro Valli$^{1,3}$
\\
$^{1}$Department of Physics and Astronomy, University of California, Irvine, CA 92697 USA\\
$^{2}$Physics Department, Yale Center for Astronomy and Astrophysics, New Haven, CT 06520, USA\\
$^{3}$ C.N. Yang Institute for Theoretical Physics, Stony Brook University, Stony Brook, NY 11794, USA}
\date{Accepted XXX. Received YYY; in original form ZZZ}

\pubyear{2022}

\begin{document}
\label{firstpage}
\pagerange{\pageref{firstpage}--\pageref{lastpage}}
\maketitle

\begin{abstract}
We explore the properties of Milky Way subhalos in self-interacting dark matter models for moderate cross sections of 1 to 5 cm$^2$g$^{-1}$ using high-resolution zoom-in N-body simulations. We include the gravitational potential of a baryonic disk and bulge matched to the Milky Way, which is critical for getting accurate predictions. 
The predicted number and distribution of subhalos within the host halo are similar for 1 and 5 cm$^2$g$^{-1}$ models, and they agree with observations of Milky Way satellite galaxies only if subhalos with peak circular velocity over all time > 7 km/s are able to form galaxies. We do not find distinctive signatures in the pericenter distribution of the subhalos that could help distinguish the models. Using an analytic model to extend the simulation results, we are able to show that subhalos in models with cross sections between 1 and 5 cm$^2$g$^{-1}$ are not dense enough to match the densest ultra-faint and classical dwarf spheroidal galaxies in the Milky Way. This motivates exploring velocity-dependent cross sections with values larger than 5 cm$^2$g$^{-1}$ at the velocities relevant for the satellites such that core collapse would occur in some of the ultra-faint and classical dwarf spheroidals.

\end{abstract}




\section{Introduction}



Self-interacting dark matter (SIDM) \citep{Spergel:1999mh, Firmani2000} is a compelling idea because it could solve the small-scale structure formation problems \citep{Kaplinghat:2015aga, Bullock:2017xww} and it arises generically in new physics models with dark sectors~\citep{Feng:2009, Loeb:2011, Tulin:2018}.

It is well established that self-interaction cross section over mass, $\sigma /m$, greater than about $1~$\cmg is required in order to impact halo densities in galaxies (see e.g.~\citet{Spergel:1999mh, Dave:2001, Colin:2002, Vogelsberger:2012ku, Rocha:2012jg, Zavala:2012us, Vogelsberger2014, Fitts2019}). When looking at spiral galaxies from the Spitzer Photometry and Accurate Rotation Curves (SPARC) data set~\citep{Lelli:2016}, \citet{Ren:2018jpt} find that SIDM can explain the diversity in rotation curves for $\sigma /m > 3$ \cmg utilizing the scatter in the concentration-mass relation for halos and the coupling between baryons and dark matter due to thermalization~\citep{Kamada:2016euw}. The strong constraints from galaxy clusters then imply that the SIDM model must have a velocity-dependent cross section~\citep{Kaplinghat:2015aga}.

Milky Way (MW) dwarf spheroidal galaxies (dSphs) play a critical role in testing the SIDM scenario because of the wide range of stellar and dark matter (DM) densities observed in these dSphs~\citep{MNRAS:Simon2019}. The predictions for the brightest of the dSphs have been intensely investigated because of the too-big-to-fail (TBTF) problem~\citep{BoylanKolchin:2011de,BoylanKolchin:2011dk}. However, it has become clear from recent investigations that lowering the densities through core expansion with a constant cross section is not sufficient to explain the diversity in densities measured for the classical and ultra-faint MW dSphs~\citep{Valli:2017ktb,Read:2018pft,Kim:2021zzw,Ebisu:2021bjh}. This has prompted investigation into larger cross section values at velocities relevant for MW dSphs such that some of the subhalos would be in the core collapse phase~\citep{Nishikawa:2019lsc,Kaplinghat:2019svz,Zavala2019,Kahlhoefer:2019oyt,Sameie:2019zfo,Correa:2021,Turner:2021,Jiang:2021foz, Sameie_2021}. 

One of the missing elements in the predictions above is the impact of the MW's stellar disk on the dSph densities. For highly radial orbits, the disk reduces the internal densities of the subhalos~\citep{Brooks:2012vi,Kelley:2018pdy} and it brings the cold, collisionless DM (CDM) predictions for the census of the satellites closer to what is observed~\citep{Graus19}. The effects of the disk are amplified for SIDM subhalos with large cores~\citep{Sameie:2018chj}, such that a detailed comparison of SIDM models to the observed satellite densities and census is not possible without including the disk. Another missing element has been the lack of a detailed prediction for the ultra-faint satellites because their internal densities are measured at radii of $\sim 100~\rm pc$ or smaller, which is difficult to resolve. 

In this work, we provide the missing elements discussed above. We analyze N-body simulations with a disk potential to simulate the properties of the hosts of the MW satellites in models with $\sigma/m = 0, 1, 5~\rm cm^2g^{-1}$. Models with cross sections in the $1-5~$\cmg range have been shown to create large cores in dwarf and low-surface brightness galaxies and are therefore excellent model targets for studying the lower-mass MW satellites. We also show that a simple analytic model developed for field halos is able to provide upper limits on the densities of the subhalos (despite the lack of resolution). This allows us to compare the densities of the subhalos to those measured in ultra-faint dSphs. 

Our key results are that cross sections in the $1-5~$\cmg range are unlikely to be consistent with the central densities of MW dSphs. We discuss the mass function of the MW satellites and the mass threshold for galaxy formation in the context of $\sigma/m = 0, 1, 5~\rm cm^2g^{-1}$. We show that $\sigma/m = 5~$\cmg is insufficient to induce core collapse in subhalos. Thus, we arrive at the conclusion that SIDM cross sections must be larger than $5~$\cmg at the velocities relevant for the MW dSphs. In the context of velocity-dependent models (for example, Yukawa-like interaction), there is a large parameter space where this can happen~\citep{Jiang:2021foz}. 

This paper is organized as follows: The simulations that we use are explained in Section \ref{sec:Simulation}. We present our findings regarding the host halo in Section \ref{sec:hostprof}, including the subhalo distributions within the host halo. We detail the subhalo properties, focusing on their circular velocity profiles, in Section \ref{sec:subprof}. We estimate the amount of mass that subhalos lose from host halo-subhalo particle scatters in Section \ref{sec:massloss}. We explore gravothermal core collapse by matching subhalos in our two SIDM simulations and using an isothermal model in Section \ref{sec:corecollapse}. Finally, we conclude in Section \ref{sec:conclusions}.

\section{Simulations}
\label{sec:Simulation}

We analyze data from two SIDM cosmological Milky Way mass ($M_{\rm vir}=10^{12} M_\odot$) halo simulations: one with a self-interaction cross section over dark matter particle mass of $\sigma / m = 1 \rm cm^2g^{-1}$ (S1), and the other with $\sigma / m = 5 \rm cm^2g^{-1}$ (S5). In addition, we compare both SIDM models with their corresponding MW halo in a CDM cosmology. The S1 simulation was previously introduced and explored in \citet{Robles:2019mfq}. The S5 simulation is the same halo as in S1 but with higher cross section. Here we provide the summary of our simulated halos (see \cite{Robles:2019mfq} for the simulation details). The MW halos were simulated using the zoom-in technique \citep{Katz_1993ApJ...412..455K,Onorbe_2013} so that the high-resolution region (with dark matter particle mass $m_{\rm DM}\approx 3\times 10^4 M_\odot$ and physical Plummer equivalent softening  $\epsilon$= 37 pc) encloses a single MW mass halo within $\sim$3 Mpc from the selected halo at $z=0$. The high-resolution region is part of a cosmological box of length  $50$ Mpc $h^{-1}$ in a Plank cosmology \citep{plank15} with $h =$ 0.675, $\Omega_m =$ 0.3121, $\Omega_b$ = 0.0488, and $\Omega_{\Lambda} = $ 0.6879. The simulations are run using GIZMO \footnote{ \url{https://www.tapir.caltech.edu/~phopkins/Site/GIZMO}} \citep{hopkins15}, and we use the SIDM implementation in \cite{Rocha:2012jg,Robles:2019mfq}. Halos are identified using \texttt{Rockstar} \citep{behroozi13a} and we use the halo catalogs that we build from the merger trees using \texttt{consistent-trees} \citep{behroozi13b}.

Simulations start at $z=125$ from the same initial conditions generated using MUSIC \citep{hahn11}. The SIDM runs are indistinguishable until $z=3$. Dark matter self interactions are modeled as velocity independent, isotropic, hard-sphere scattering with cross section $\sigma$. Velocity dependence and angular dependence likely have measurable effects on halo and subhalo evolution, which is an interesting direction to explore in the future.

Accounting for the baryonic matter is necessary to infer more realistic predictions for MW satellites. In this work, we account for the gravitational effects of baryonic matter by including disk and bulge potentials following the semi-analytical model in \cite{Kelley:2018pdy}, initialized at $z$=3. Hereafter, we will only present results for the CDM and SIDM simulations that include the contribution of the baryonic potentials (unless specified otherwise).

The initial conditions for the simulations, including the baryonic potentials, are described in \citet{Kelley:2018pdy} for CDM, and \cite{Robles:2019mfq} for SIDM. For the S5 simulation that we introduce in this work, we use the same initial condition as S1 which includes the baryonic potential at $z$=3. The bulge component is modeled by a Hernquist sphere. The disk and bulge (which we will refer to as just "disk" throughout this paper) are set to agree with observations at $z$=0. The disk is composed of stars and gas with a stellar radius of $R_*$ = 2.5 kpc, stellar height $h_*$ = 0.35 kpc, stellar mass $M_*$ = $4.1 \times 10^{10} M_\odot$, gas radius $R_{\rm gas}$ = 7 kpc, gas height $h_{\rm gas}$ = 0.084 kpc, and gas mass $M_{\rm gas}$ = $1.86 \times 10^{10} M_{\odot}$ at $z$=0.

This work is focused on one simulated halo with the baryonic disk and bulge tuned to the MW, so running different halo masses would be useful. The effect of the baryonic potential on host halo shape \citep{Vargya_2022} and star formation rate \citep{Sameie_2021} have been explored in SIDM. In a CDM scenario, \citet{Garrison_Kimmel_2017} find that subhalo mass fraction and radial distribution are minimally effected by disk radius, thickness, and shape. Increasing the disk mass by a factor of two only decreases the number of subhalos within $50$ kpc by about $20$\% \citep{Garrison_Kimmel_2017}.
While this effect may be slightly amplified in an SIDM scenario, we don't expect disk parameters to effect our results, unless disk mass is varied by a factor of at least two.

\section{Host Halo}
\label{sec:hostprof} 

In this section we discuss the effects that SIDM has on the host halo. We have chosen to look at two observable quantities: the particle distribution within the host halo and the subhalo distribution within the host halo.

\begin{figure*}
	\includegraphics[width=\textwidth]{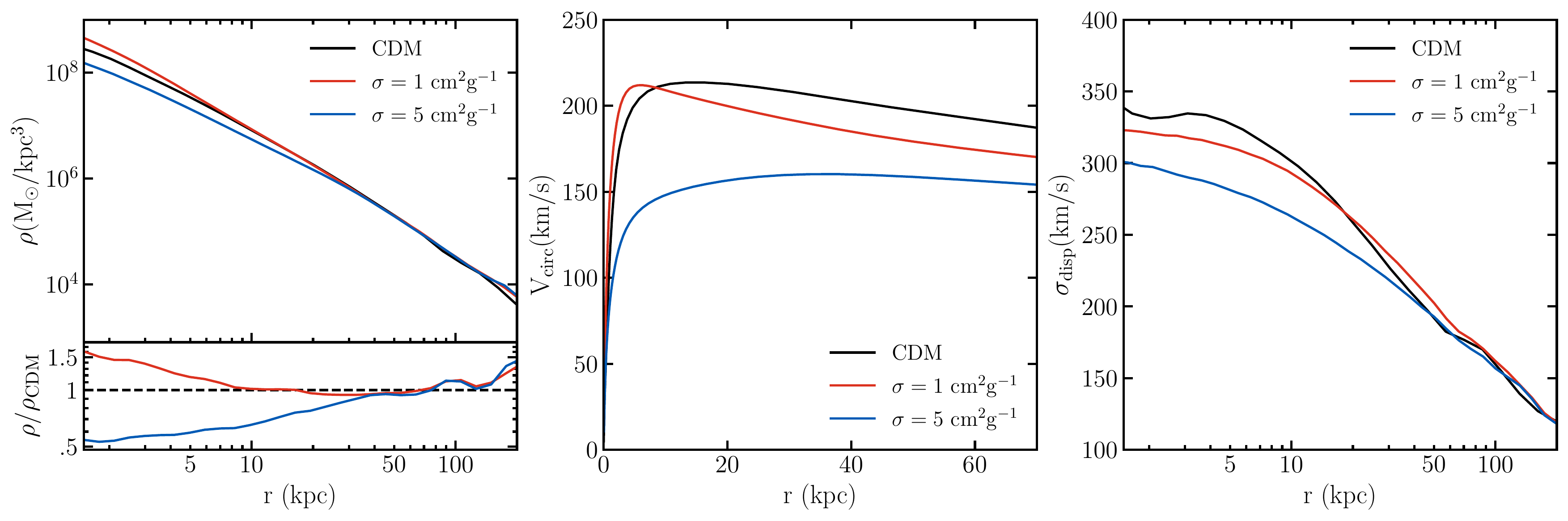}
    \caption{Host halo properties at $z$=0 for CDM (black), $\sigma/m = 1$ cm$^2$g$^{-1}$ (red), and $\sigma/m = 5$ cm$^2$g$^{-1}$ (blue). Left: Dark matter density profile. All three models show similar densities outside 20 kpc and are cuspy in the inner regions. Middle: Dark matter circular velocity profile. The difference in maximum circular velocity attests to the lower central density in S5. Right: Dark mater velocity dispersion. S1 and S5 have thermalized central regions, while CDM rises within the stellar and gas disk ($R_* = 2.5$ kpc at z=0). S5 has a lower density and velocity dispersion within about 20 kpc because of the thermalization process expected with SIDM.}
    \label{fig:hostden}
\end{figure*}

\subsection{Particle Distribution}

First, we compare the density profile ($\rho$), the circular velocity ($V_{\rm{circ}}$), and velocity dispersion ($\sigma_{\rm disp}$) of the host halo in the CDM and SIDM simulations. We use particle data from each simulation to calculate the current time ($z$=0) profiles. In Appendix \ref{sec:apphostprof}, we additionally solve for the Navarro-Frenk-White (NFW)~\citep{Navarro:1996gj} density profiles at very early times ($z$=3) and at $z$=0. We find that there is negligible difference between the NFW early and late time profile and the particle density profile in the region $r > 10$ kpc. Since the subhalos we investigate in this paper always stay outside 10 kpc, we use the $z$=0 particle density of the host halo throughout this work.

Figure \ref{fig:hostden} shows the DM density profiles of the host halo in CDM, S1, and S5 in the left panel, the circular velocity profiles in the center panel, and the velocity dispersion profiles in the right panel.
Density and velocity dispersion are calculated within 40 logarithmically spaced radial bins. Velocity dispersion is calculated as $\sigma_{\rm disp} = \sqrt{\sigma_{\rm{x}}^2 + \sigma_{\rm{y}}^2 + \sigma_{\rm{z}}^2}$ where $\sigma_{\rm{x}}^2 = \langle (V_{\rm{x, i}} - \langle V_{\rm{x}} \rangle) ^2 \rangle$. Here $V_{\rm{x, i}}$ is the velocity in the x direction of each particle, $i$, within the radial range. Circular velocity is given by Newtonian gravity as $\sqrt{GM/r}$ where $M$ is the enclosed mass within radius $r$.

Both SIDM models follow CDM to $\sim$40\% in density outside 10 kpc and $\sim$30 \% outside 20 kpc, indicating that objects orbiting the host halo in all three simulations spend a majority of their time in areas of similar density. The density profiles of the host halos in all three models have small cores and are rather cuspy. At $r$ = 1 kpc, S5 is $\sim$ 0.5 times as dense as CDM, while S1 is $\sim$ 1.5 times as dense as CDM. 

In the absence of a baryonic disk, SIDM host halos have been shown to have larger cores and lower central densities \citep{Nadler:2020ulu}. 
Due to the dependence of the DM density on the baryon gravitational potential \citep{Kaplinghat:2013xca}, SIDM halos are expected to have smaller cores and higher densities when taking the baryonic disk into account. A comparison of S1 and CDM with and without the effects of a baryonic disk is given by \citet{Robles:2019mfq}.

The middle panel of Figure \ref{fig:hostden} shows that the maximum circular velocity of host halos in CDM and S1 are both just over 200 km/s, while the maximum circular velocity of the host halo in S5 is around 150 km/s. This difference is due to the low central density in the S5 host halo with respect to S1 and CDM.

The right panel of Figure \ref{fig:hostden} shows the velocity dispersion. In the CDM model, the velocity dispersion peaks at about 4 kpc and rises steeply within $\sim$2.5 kpc, which corresponds to the radius of the stellar and gas disk. When DM self-interactions are included, velocity dispersion corresponds to temperature. 
The about constant velocity dispersions at the center of both SIDM models indicate that these regions are isothermal. The hot core and colder outer region leads to gravothermal core collapse in the SIDM host halos, as also noted by \citet{Elbert:2018ApJ} and \citet{Robles:2019mfq}.
S5 has a slightly lower velocity dispersion at the center of the halo than S1, because with a larger interaction rate, heat transfers to the outer parts of the halo more efficiently. 

\emph{Thus, in the regions in which subhalos orbit, the host halo density profiles in the CDM, S1, and S5 models are comparable to within 30\%. The 50\% difference in the host central densities leads to slightly different circular velocity profiles in all three models. The host halos in both SIDM simulations are isothermal, while the CDM halo is not.}

\subsection{Subhalo Distribution}
\label{sec:subdist}

\begin{figure*}
	\includegraphics[width=\textwidth]{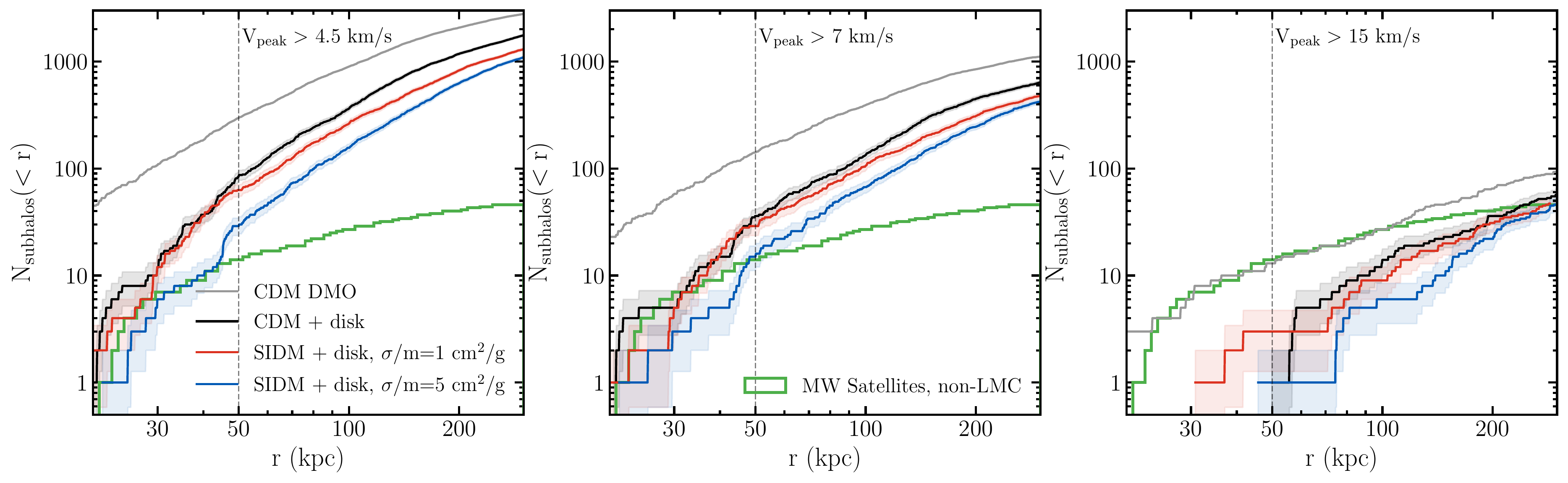}
    \caption{Cumulative number of subhalos within the virial radius of the host halo at $z$=0. We compare CDM dark matter only (gray), CDM with the baryonic disk (black), S1 with the baryonic disk (red), and S5 with the baryonic disk (blue) to observed Milky Way satellite galaxies (green). The left, middle, and right panels show subhalos with \Vpeak cuts of 4.5, 7, and 15 \kms, respectively. The vertical dashed line is a reminder that observations are likely not complete outside 50 kpc and we expect many other satellite galaxies to be observed in this region with upcoming telescopes.}
    \label{fig:subdisthist}
\end{figure*}

\begin{figure}
	\includegraphics[width=\columnwidth]{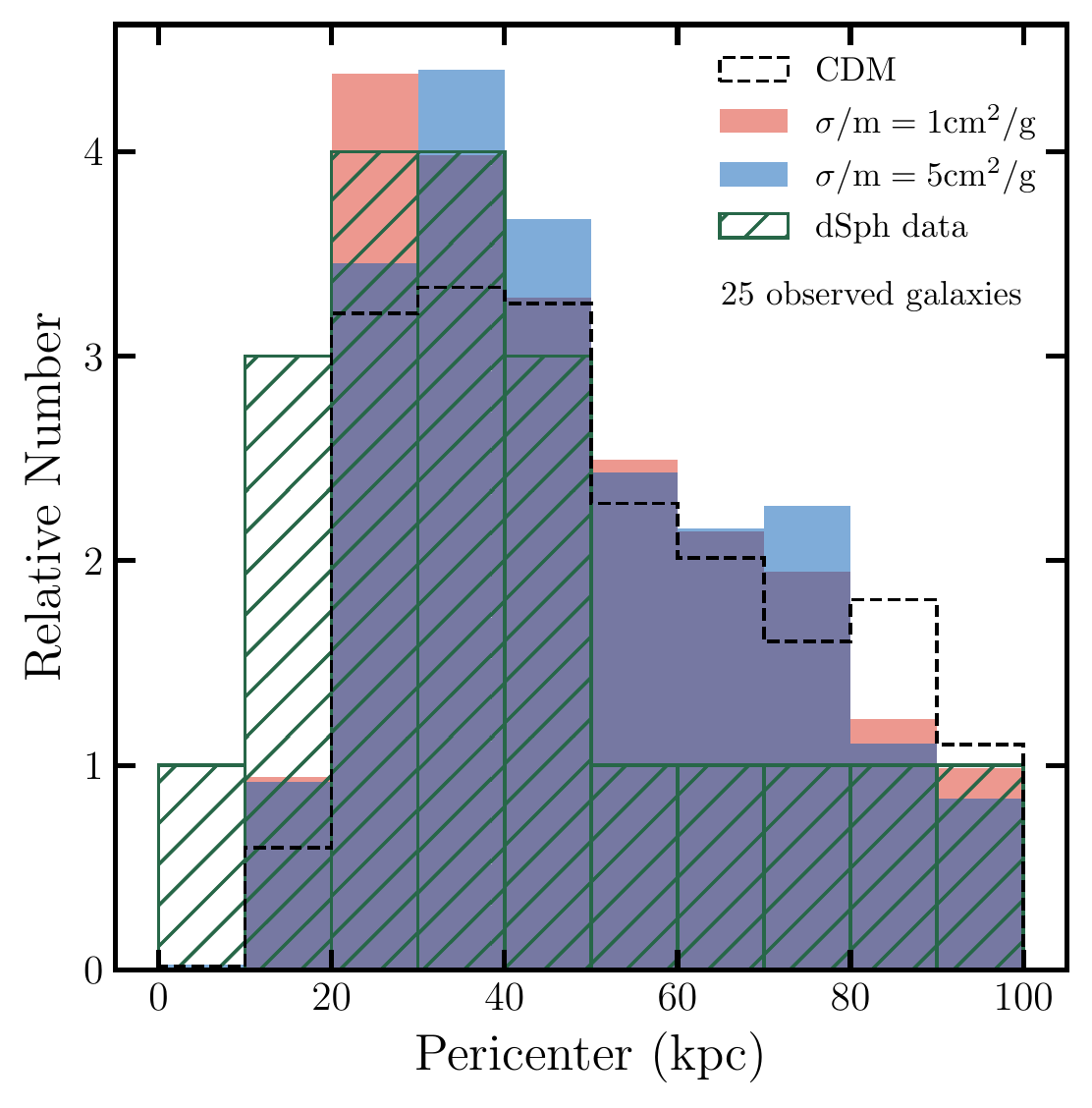}
    \caption{Pericenters of subhalos within the virial radius of the host halo and with \Vpeak > 4.5 \kms at $z$=0. The Relative Number is equal to the number of subhalos at a given pericenter over the total number of observed dSphs, such that the sum of any one histogram is equal to the number of observed galaxies (in our case 25).
    We compare CDM (black), S1 (red), S5 (blue), and observed dSphs (green). We only include dSphs that are currently within 270 kpc from the center of the galaxy. Histogram values for dSph data are found by taking the median of each bin after sampling the distribution of each dSph $10^3$ times. Our sampled dSph data have Poisson distributions, indicating that $\sqrt{\rm N}$ error bars are appropriate.}
    \label{fig:subperihist}
\end{figure}

We now focus on the subhalo distribution within the host halo. Using $z$=0 halo catalogs from merger trees, we look at the radial distribution of subhalos within 300 kpc from the center of the host halo. In our simulations, subhalos are well resolved if $V_{\rm{peak}} > 4.5$ \kms, which corresponds to a bound mass of about $5 \times 10^6 M_{\odot}$ \citep{Robles:2019mfq}. \Vpeak is the maximum $V_{\rm{max}}$ throughout the subhalo's history and $V_{\rm{max}}$ is the maximum circular velocity of the subhalo at any given time. We will look at both the current position of subhalos, and their pericenters.

\subsubsection{Current distance}

Figure \ref{fig:subdisthist} shows the number of subhalos currently within a given radius of the host halo for CDM dark matter only (DMO), CDM with the baryonic disk, S1 with the baryonic disk, and S5 with the baryonic disk. In this section, we focus on three populations of subhalos: all well resolved subhalos ($V_{\rm{peak}} > 4.5$ \kms), more massive subhalos ($V_{\rm{peak}} > 7$ \kms), and the most massive subhalos ($V_{\rm{peak}} > 15$ \kms). Each panel shows subhalos with a different \Vpeak cut. To compensate for the variance among simulations we estimate an error of $\sqrt{N_{\rm{subhalos}}}$ for the CDM, S1, and S5 simulations. To compare the simulations to observations, we plotted the radial distribution of the observed MW satellites compiled in \citet{MNRAS:Simon2019}, excluding the six satellites that \citet{Patel_2020} find are likely to be satellites of the Large Magellanic Cloud (LMC). We do not include any sky correction for satellites yet to be found, so the plotted curve should be treated as a lower limit to the actual radial distribution. 

When looking at all resolved subhalos (\Vpeak > 4.5 km/s) at and outside 50 kpc, all models produce more subhalos than observed. With the sensitivity and breadth of upcoming telescopes like the Vera C. Rubin Observatory, more satellite galaxies will likely be found, especially outside $\sim$50 kpc. We do not focus on the radial distribution of satellites below 50 kpc because this region is more sensitive to the LMC.
When looking at more massive subhalos (\Vpeak > 7 km/s), S5 produces the same number of subhalos as observed galaxies within 50 kpc. CDM and S1 slightly over-predict the number of subhalos within 50 kpc, and we find that for only a slightly larger value of \Vpeak > 8 km/s, CDM and S1 also produce the same number of subhalos as observed galaxies within 50 kpc.
Again, all simulations over-predict the number of galaxies outside 50 kpc, which is likely due to incompleteness. Finally, when looking at only the most massive subhalos (\Vpeak > 15 km/s), CDM, S1, and S5 under-predict the number of observed galaxies.
Only CDM DMO (which is clearly not representative of the MW) has about the same number of subhalos as the number of observed galaxies within 50 kpc. 
Thus our simulations predict that \Vpeak must be below 15 \kms to be consistent with the observed galaxy sample, irrespective of the dark matter model. 

It has been shown that not all subhalos host luminous galaxies \citep{Sawala_2015, Simpson_2018, Benson:2001at, Somerville_2002, Bullock_2000}, and the fraction of subhalos that do host luminous galaxies, depends on subhalo mass, reionization redshift, and baryon survival criterion \citep{Kim_2018} of the model. There have been many attempts to model the luminosity fraction, and for CDM DMO scenarios, a criteria of $V_{\rm{peak}} > 23.5$ \kms is required to match simulations to observations \citep{Dooley_2017}. This limit shifts to $V_{\rm{peak}} > 7$ \kms for CDM + disk scenarios \citep{Graus19,Kelley:2018pdy}. If we assume that the observed satellite data are complete within 50 kpc, S1 and S5 halos with $V_{\rm{peak}}$ down to $7$ \kms are required to match the total number of observed satellites within 50 kpc, as seen in Figure \ref{fig:subdisthist}. 

This comparison between simulated subhalos and observed satellite galaxies within 50 kpc illuminates the trend that both CDM and SIDM models struggle to produce observed subhalo numbers at larger \Vpeak values. This result is based on one set of simulations, where our aim has been to compare the impact of dark matter  self-interactions on this issue. A more through statistical analysis with a large set of simulations is clearly warranted given the importance of this issue. 

For CDM, a Bayesian comparison between observed satellites and simulated subhalos at all radii by \citet{Jethwa_2017} found that simulations match observations if subhalos with peak virial mass $M_{\rm vir} > 2.4 \times 10^8 M_{\odot}$ (corresponding to $V_{\rm{peak}} > 15$ km/s) form the faintest MW satellites. This is at odds with the results plotted in Figure \ref{fig:subdisthist} given the stark differences in the observed vs simulated satellites for  \Vpeak > 15 km/s. The reasons for these large differences are not clear, and this should be studied further. We note that the largest effect on the subhalo population within $50$ kpc (in our simulations) is due to the disk of the MW. For the disk mass and disk radius assumed here, the number of surviving CDM subhalos within $50$ kpc is only about $30$\% of the number in the corresponding DMO simulation. What this implies is that the surviving satellite population is likely to be very sensitive to the disk modeling.

Some of this subhalo destruction may be due to artificial disruption. When comparing CDM N-body simulation, \emph{Bolshoi} \citep{Klypin_2011}, to \texttt{SatGen} \citep{Jiang_2021_satgen}, a semi-analytical subhalo evolution modeling framework  which can track orphan galaxies, \citet{Green_2021} find that accounting for orphan galaxies can boost the number of subhalos by up to a factor of two in the inner regions. With a factor of two more subhalos within $50$ kpc, S5 would match the number of observed satellites for \Vpeak values greater than at least $7$ km/s. However, \citet{Green_2021} predict only 0.1 dex more subhalos below a mass of $m_{\rm sub} / M_{\rm host} < 10^{-4}$. Thus, artificial subhalo disruption should not greatly affect the results presented in Section~\ref{sec:subprof} and onward.

A further complication is that, while we have removed some satellites that could have been associated with the LMC, we haven't included the gravitational impact of the LMC on the other satellites and the survival probability of the satellites in a consistent manner. This point has been noted in the literature \citep{Nadler_2021, Nadler_2020, Bose_2020}. It is critical to explore this in more detail if we are to use the MW satellites to put a lower limit on the minimum halo mass to form galaxies. In the present analysis, we therefore focus our attention on the count within 50 kpc and not the detailed radial distribution, which is likely to have been impacted by the LMC.

These comments and results connect to  interesting questions in galaxy formation at the faint end.  Efforts in this regard have found that only dark matter halos with $V_{\rm{peak}} > 20$ \kms form stars \citep{Thoul_1996, Okamoto_2008, Ocvirk_2016, Fitts:2016usl}. Subhalos with \Vpeak = 7 \kms have a virial temperature of about 2000 K \citep{Graus19}, well below the atomic hydrogen cooling limit, $10^4$ K. This suggests that in our S1 and S5 simulations (even when orphan galaxies are accounted for), as well as in CDM+disk simulations, subhalos would require molecular cooling to form stars \citep{Graus19,Kelley:2018pdy}.

\citet{Manwadkar_2021} come to a similar conclusion using the Caterpillar CDM simulation suite, which has comparable resolution to the simulations presented in this work but does not account for the gravitational potential of the baryonic disk. They are able to reproduce the observed radial distribution of satellite galaxies (when not accounting for orphan galaxies) only if halos below the atomic cooling limit are able to form stars.

Thus, it seems that the Missing Satellites (MS) problem has drastically changed since its original assessment \citep{Kim:2021zzw}. Initially, the number of simulated subhalos far outnumbered observed satellite galaxies of the MW \citep{Kauffmann_White_Guiderdoni_1993, Klypin:1999uc, Moore:1999nt, DOnghia:2003}. At the time, simulations only included CDM-only models. Since then, more developed simulations that include baryonic matter as well as novel DM models have been studied. Observational tools have also improved, allowing old, faint, and low surface brightness satellites to be found~\citep{McConnachie:2012vd, Fritz:2018aap, Drlica-Wagner:2015ufc, Drlica-Wagner:2020}. With these advancements, the MS problem has almost been turned on its head: potentially simulating far fewer subhalos than there are observed satellite galaxies \citep{Kim_2018,Graus19,Kelley:2018pdy}, as can be seen in the right panel of Figure \ref{fig:subdisthist}.

\emph{In order for the number of subhalos produced in S1 and S5 simulations to agree with observations within 50 kpc, we find that stars must form in subhalos with \Vpeak as low as 7 \kms. If only halos with $V_{\rm{peak}} > 20$ \kms can form stars, as suggested by some galaxy formation simulations, CDM DMO, CDM + disk, S1, and S5 all under-predict the number of observed satellites.}

\subsubsection{Subhalo pericenters}
\label{sec:subperi}
We also look at the pericenter distributions of subhalos in the CDM, S1, and S5 simulations. Pericenter is defined as the minimum distance to the center of the host halo over all time. As noted by \citet{Garrison_Kimmel_2017} a subhalo's orbit can drastically affect it's survival probability when the baryonic disk is taken into account. More specifically, few subhalos that come within 10 kpc of the center of the host halo survive. 

Figure \ref{fig:subperihist} shows pericenter histograms for CDM, S1, and S5 compared to \emph{Gaia} data of satellite galaxies. In green, we show the satellite galaxies currently within 270 kpc of the host halo. The dashed black, solid red, and solid blue histograms show pericenters of simulated subhalos in CDM, S1, and S5 that are currently within 270 kpc of the host halo and have \Vpeak > 4.5 \kms. The simulated subhalos are normalized to the number of observed halos for easier comparison.

For the CDM subhalos, pericenter is calculated by interpolating between snapshots (spaced by 14-16 Myr) and selecting the minimum distance to the host \citep{Kelley:2018pdy}. For the SIDM simulations, we calculate pericenter by conserving energy and angular momentum between the two snapshots that straddle the subhalo's closest approach. This simple calculation, detailed in Appendix \ref{app:pericalc}, assumes that the gravitational potential of the host halo is spherical and that no angular momentum is lost to dynamical friction. For cases where these assumptions do not hold we simply take the pericenter to be the minimum distance to the host captured in the simulation snapshots. To ensure that our calculated pericenters are reasonable, we only use calculated pericenter values that vary from the minimum distance captured in the simulation snapshots by less than 50\%.

To compare simulation to observation, we use the inferred pericenter of satellite galaxies with respect to the MW (with a mass of $1 \times 10^{12} M_{\odot}$), taking into account the LMC with a mass of $1.8 \times 10^{11} M_{\odot}$ \citep{Patel_2020}. We note that \citet{Patel_2020} study all satellite galaxies that are associated with the LMC based on their membership to the MW's Vast Polar Structure. For satellites not given by \citet{Patel_2020}, we use pericenter data with respect to the MW (with a mass of $0.8 \times 10^{12} M_{\odot}$) assuming no effects by the LMC \citep{Fritz:2018aap}. Using only \citet{Fritz:2018aap} data, we found a similar histogram but with slightly smaller pericenters. Using a heavier MW mass ($1.6 \times 10^{12} M_{\odot}$), we find a histogram with even smaller pericenters, as expected. We chose to show the lighter MW mass because it is more consistent with our simulated MW mass ($1 \times 10^{12} M_{\odot}$). To account for the errors in inferred pericenters, we sampled the distribution of each dSph $10^3$ times and plotted the median value for each pericenter bin.

Figure \ref{fig:subperihist} shows that all three DM models predict a similar histogram to that of the observed dSphs, peaking at 20-40 kpc. We do notice a slight discrepancy between simulation and observation at pericenters below 20 kpc and at about 60 kpc. In each of these regions, we find that the sampling of the data gives a Poisson distribution, implying error bars of $\sqrt{N}$. While the discrepancy under 20 kpc was also noted in \citet{Kelley:2018pdy}, we find it has been slightly alleviated when accounting for the LMC.~\footnote{The predicted pericenter distribution peaks below 10 kpc in simulations that do not include central disk potentials \citep{Kelley:2018pdy}, in significant disagreement with the observed distribution.} The remaining discrepancy within 20 kpc may be due to the non-identification of tidally and artificially disrupted halos that fall below the \texttt{Rockstar} resolution limit \citep{Green_2021}. The discrepancy at about 60 kpc may be due to incomplete observations of the faintest dSphs outside 50 kpc. The middle panel of Figure \ref{fig:subdisthist} shows that there may be about 300 dSphs within 270 kpc rather than the 25 observed satellites that we include in Figure \ref{fig:subperihist}. These dSphs, which may be observed by upcoming telescopes like the Vera C. Rubin Observatory, would dominate the relative number count at larger pericenters.

Since the projection of dSphs pericenters does change based on the MW and LMC model, a range of MW and LMC models would be needed to make a more robust conclusion here. It would be interesting to study how the MW and LMC models would need to change in order to match the observed and simulated pericenter distributions. Our results indicate that pericenter distributions by themselves cannot distinguish between different cross sections in the 0-5 \cmg range.

Looking at subhalo pericenters in conjunction with other observable quantities may prove useful in the future~\citep{Correa:2021}. A correlation between pericenter and the density at 150 pc is found among observed dSphs by \citet{Kaplinghat:2019svz}. We looked at pericenter in comparison to the density at 200 pc, and we found no correlation for subhalos in our S1 and S5 simulations. \citet{Robles_2021} found a correlation between pericenter and $V_{\rm{max}}$ and $r_{\rm{max}}$. This correlation is discussed further in Section \ref{sec:subprof}. 

\emph{We are not able to use pericenter distributions alone to differentiate between CDM, S1, and S5, and we find a slight discrepancy between all three DM models and observations in subhalos with pericenters less than 20 kpc.}

\section{Subhalo Profiles}
\label{sec:subprof}

\begin{figure*}
	\includegraphics[width=\textwidth]{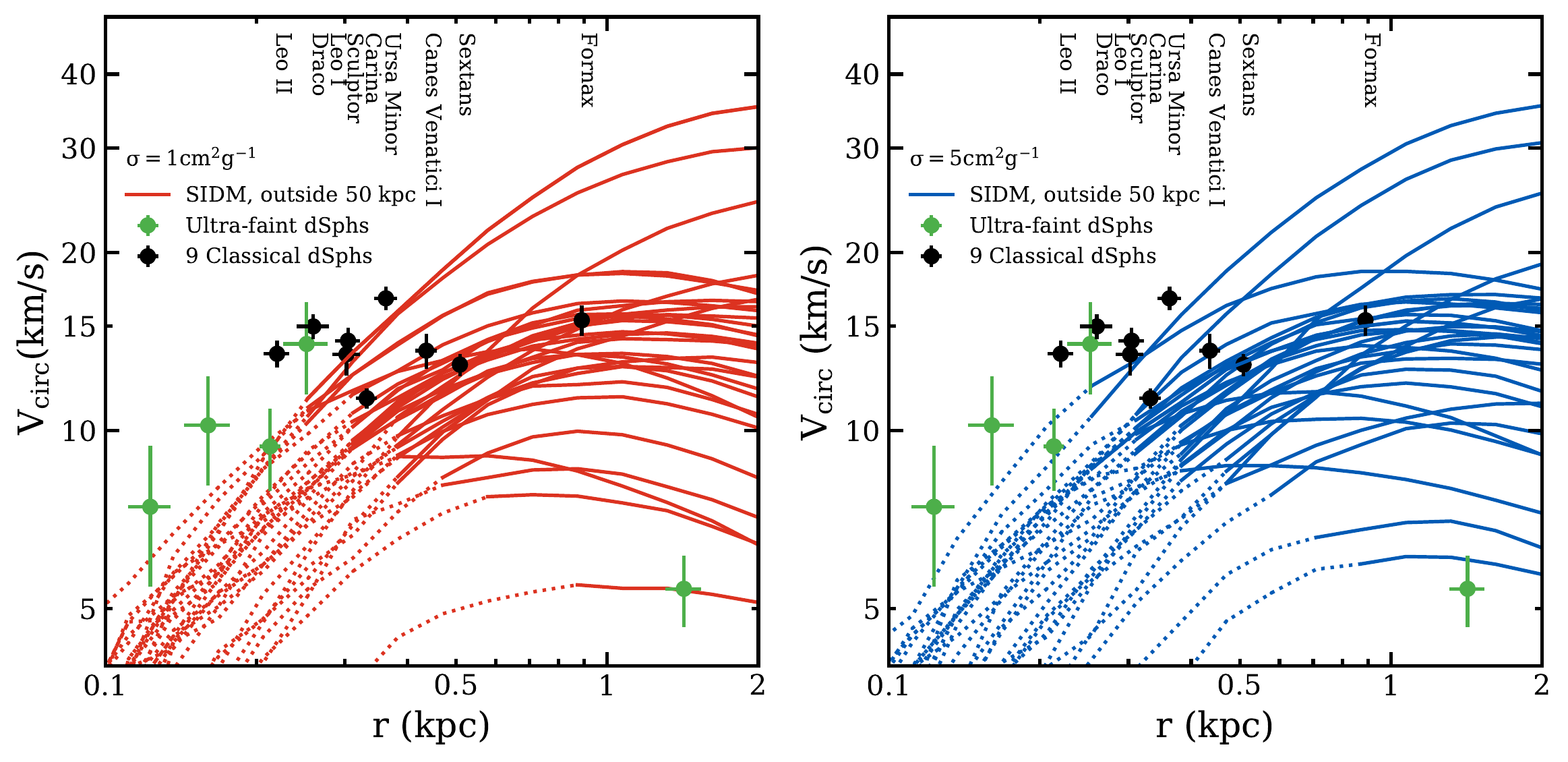}
    \caption{Circular velocity of the 30 subhalos that reside between 50 kpc and the host halo virial radius at $z$=0 with largest \Vpeak that are resolved at $r_{\rm{max}}$. We compare both S1 (left) and S5 (right) to the nine classical dwarf spheroidal (dSphs) MW satellite galaxies (black), which all reside outside 50 kpc from the MW center. We also compare to ultra-faint dSphs (green), some of which reside within 50 kpc and some which reside outside 50 kpc. For all dSphs, we plot the sphericallized half-light radius, $r_{1/2}$, and velocity at half-light radius, $V_{1/2}$. The dotted lines correspond to regions where the subhalos are not resolved, containing less than 200 particles.}
    \label{fig:vcircover}
\end{figure*}

\begin{figure*}
	\includegraphics[width=\textwidth]{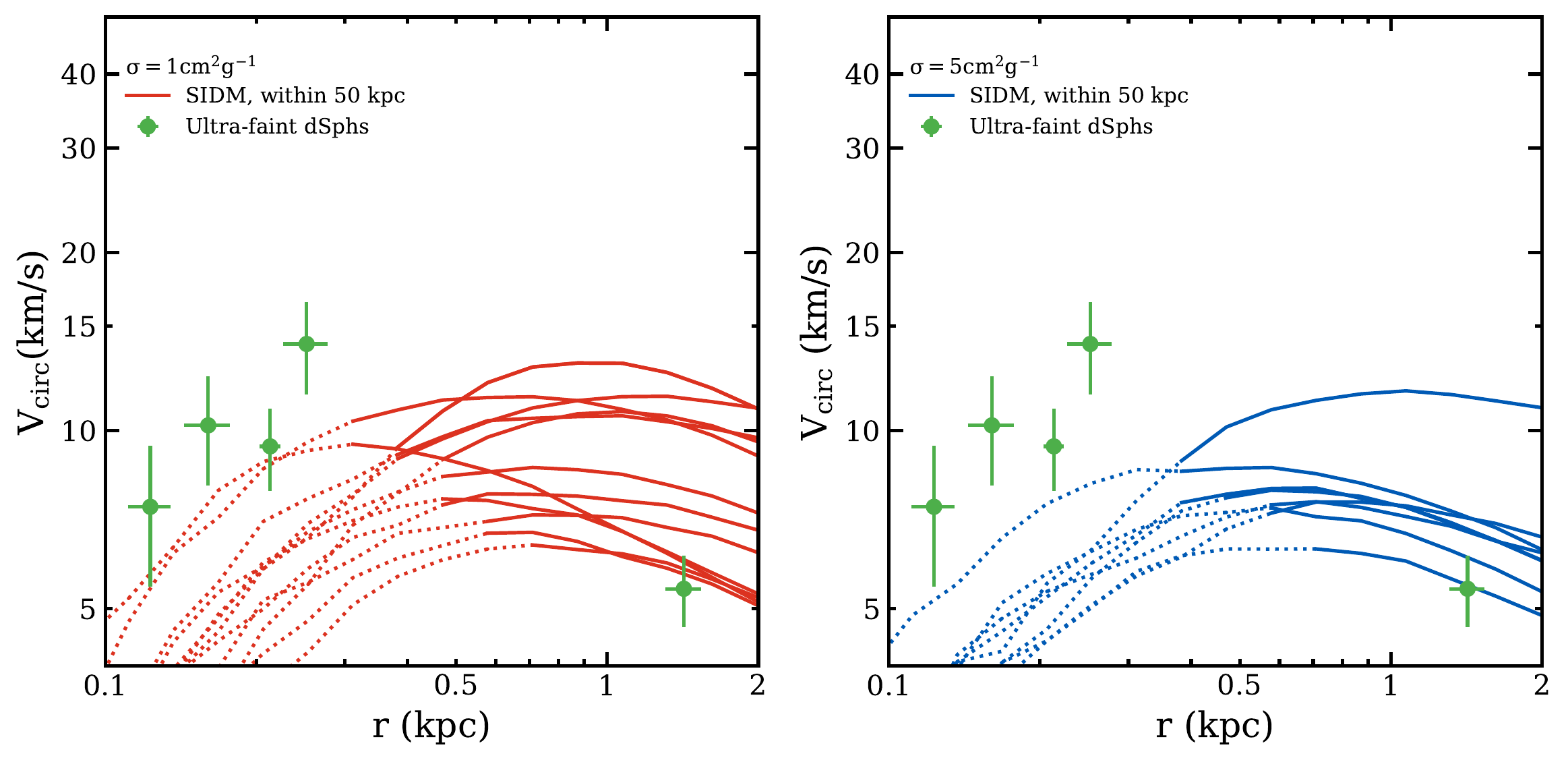}
    \caption{Circular velocity of all subhalos that reside within 50 kpc of the host halo center at $z$=0 that are resolved at $r_{\rm{max}}$. We have 12 resolved S1 subhalos and 8 resolved S5 subhalos. We compare to the sphericallized $r_{1/2}$ and $V_{1/2}$ of ultra-faint dSphs.}
    \label{fig:vcircunder}
\end{figure*}

\begin{figure}
    \centering
    \includegraphics[width=\columnwidth]{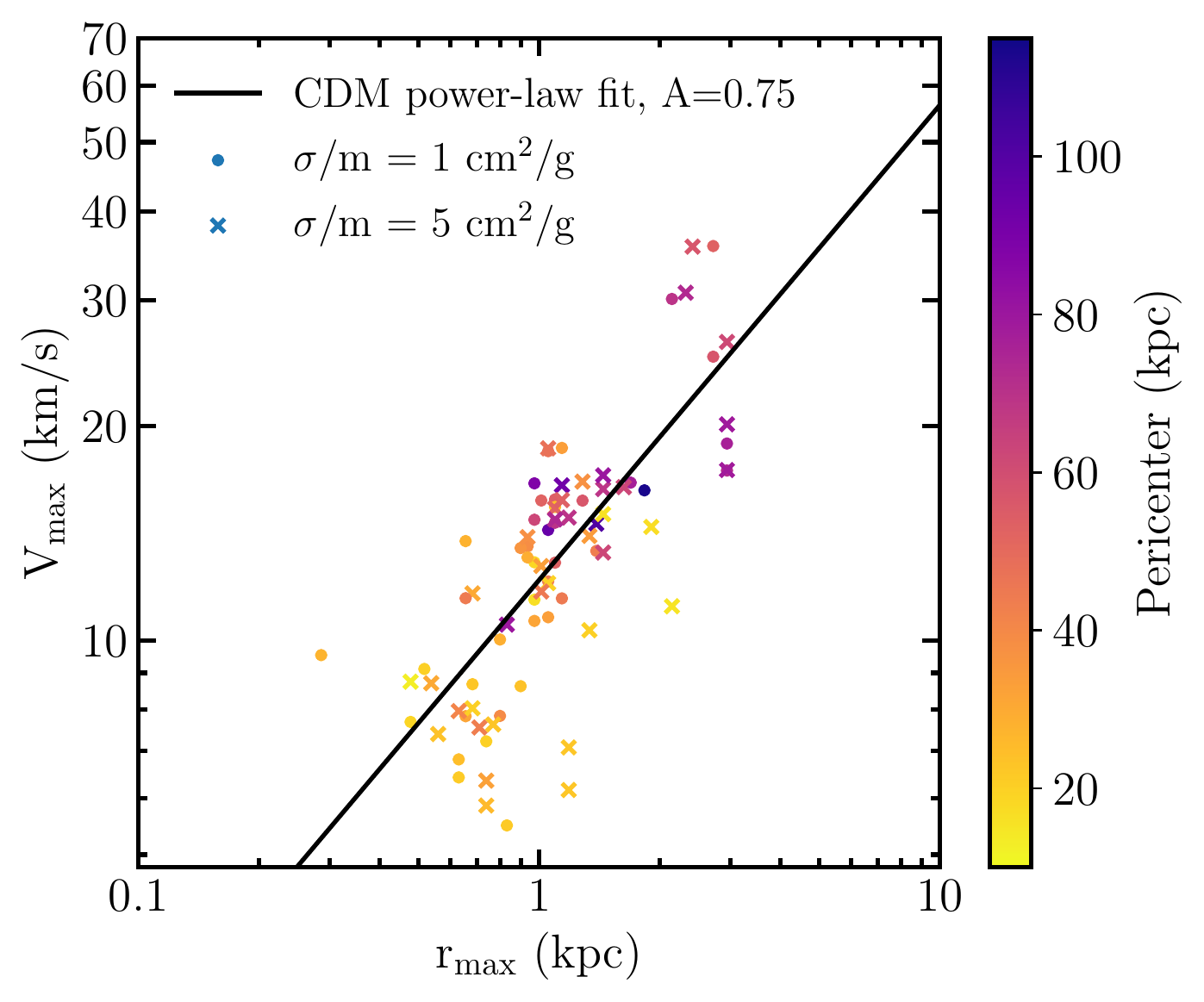}
    \caption{The maximum velocity as a function of the radius at maximum velocity at z=0 of the 30 resolved subhalos outside 50 kpc with largest \Vpeak and all resolved subhalos within 50 kpc. The dots and x's represent subhalos in the S1 or S5 simulation, and the color is determined by the pericenter. The black line shows the power-law fit for subhalos in the CDM+disk simulation \citep{Robles_2021}.}
    \label{fig:VmaxRmax}
\end{figure}

In this section we look at the z=0 particle distribution within subhalos in the SIDM simulations. We focus on their circular velocity profiles, $V_{\rm{max}}$, and $r_{\rm{max}}$ values, but also show their density profiles in Appendix \ref{sec:appdenprofs}. We define $r_{\rm{max}}$ as the radius at which $V_{\rm{max}}$ occurs. To make sure we can trust our $V_{\rm{max}}$ and $r_{\rm{max}}$ values, we only look at subhalos that are resolved at $r_{\rm max}$, which, from convergence studies \citep{Power03,Hopkins18} amounts to having $\geq$ 200 particles within $r_{\rm{max}}$. We use this cut because subhalos with fewer particles may undergo significant artificial stripping in the simulation. We use this resolution cut in addition to the resolution cut specified in Section \ref{sec:subdist}: \Vpeak > 4.5 \kms. 

Using these resolved subhalos, we further define two populations. The first population of subhalos are currently within 50 kpc of the host center. These inner subhalos have come relatively close to the center of the host halo.
The second population of subhalos are currently outside 50 kpc but within the virial radius of the host halo, $R_{\rm{vir, host}}$. For all simulations that we use in this paper, $R_{\rm{vir, host}} \approx 270$ kpc. These outer subhalos have likely not come as close to the center of the host halo but are likely bound to the host halo.

We use 50 kpc to separate the two populations to distinguish subhalos that have been through more dense regions of the host halo from those that have not. While the current position of a subhalo does not convey information about the subhalo's history, we find that it sufficiently distinguishes subhalos that have come closer to the host halo from those that have not. In section \ref{sec:massloss}, we additionally look at the subhalos' pericenters to convey information about their history.

The number of subhalos within 50 kpc with \Vpeak > 4.5 \kms and for which $r_{\rm{max}}$ encloses 200 particles is so low that we are able to look closely at all of these subhalos. In the S1 simulation, we have 12 such subhalos, and in the S5 simulation, we have 8. There are many subhalos outside 50 kpc with \Vpeak > 4.5 \kms and for which $r_{\rm{max}}$ encloses 200 particles, so we take a closer look at a subset of subhalos within the outer population that have the largest $V_{\rm peak}$, which corresponds to the most massive subhalos. Being the most massive subhalos, they most likely host massive dSphs, similar to the ones that have been observed in the MW. We look at 30 subhalos currently outside 50 kpc in each SIDM simulation. We chose to look at 30 subhalos to roughly replicate the number of observed dSphs.

To look at subhalo characteristics without influence from host halo particles, we distinguish subhalo and host halo particles from one another using their velocity with respect to the velocity of the subhalo, which we denote as the relative velocity. Particles that are bound in the subhalo have relative velocities $\mathcal{O}$(10 \kms), capping at about the maximum circular velocity, $V_{\rm{max}}$, of the subhalo. Subhalos are orbiting the host halo at a velocity $\mathcal{O}$(100 \kms), so the relative velocity of the host halo particles will be much larger than the relative velocity of the subhalo particles. When looking at subhalo characteristics, we cut out any particles with $V > 3 V_{\rm{max, sub}}$, leaving only the subhalo particles. Assuming particles in the subhalo have negligible velocity dispersion, cutting $V > V_{\rm{max, sub}}$ removes all subhalo particles. We cut $V> 3 V_{\rm{max, sub}}$ to account for velocity dispersion. The current circular velocity, $V_{\rm{circ}}$, as a function of radius of all of the selected subhalos are shown in Figures \ref{fig:vcircover} and \ref{fig:vcircunder}. 

We also show two populations of observed dSphs, 1) the nine classical dSphs, and 2) a subset of ultra-faint dSphs with the cleanest data \citep{Kaplinghat:2019svz}: Ursa Major I, Ursa Major II, Bo{\"o}tes I, Hercule, Coma Berenices, Canes Venatici II, Crater II, Reticulum II, Segue I, and Antlia II. Antilia II is a newly discovered and interesting satellite to compare to SIDM because it is the most diffuse satellite yet observed \citep{Sameie_2020}. Note that not all ultra-faint dSphs are shown in Figure \ref{fig:vcircover} because of the radial range shown.

To compare the dSphs to the subhalo circular velocity profiles, we plot the inferred dSph circular velocity at half-light radius, $V_{1/2}$, versus the 3-Dimensional deprojected half-light radius, $r_{1/2}$. We calculate $V_{1/2}$ using \citep{Wolf:2009tu}:
\begin{equation}
\label{eq:v12}
    V_{1/2} = \sqrt{\frac{GM_{1/2}}{r_{1/2}}} = \sqrt{3\langle \sigma_{\rm{los}} \rangle ^2}
\end{equation}
where $M_{1/2} = \rho_{1/2} (4/3 \pi r_{1/2}^3)$ and $\langle \sigma_{\rm{los}} \rangle$ is the average line of sight velocity dispersion. 

For all dSphs, we use the 2-Dimensional half-light radius, $R_{\rm{e}}$, from \citet{MNRAS:Simon2019} (except for Antlia II, which we get from \citet{Torrealba:2019}) and convert to $r_{1/2}$ using the conversion factor of 4/3 \citep{Wolf:2009tu}. We also modify $r_{1/2}$ by a factor of $\sqrt{1- \epsilon}$, where $\epsilon$ is the ellipticity given in \citet{MNRAS:Munoz2018}, to assume spherical symmetry for the dSphs, just as we do for our simulated halos (we use this sphericallized radius in Figures \ref{fig:vcircover} and \ref{fig:vcircunder}). In Appendix \ref{app:nosph} we show the same dSphs \textit{without} assuming spherical symmetry. For the nine classical dSphs we take $\rho_{1/2}$ from the Jeans analysis in \citet{Kaplinghat:2019svz}. For the ultra-faint dSphs, we use the mass estimator in \citet{Wolf:2009tu} with the data in \citet{MNRAS:Simon2019}. The stellar mass contributions to $V_{\rm 1/2}$ are negligible for all dSphs. In the most extreme case of Fornax, stellar mass only contributes to $V_{\rm 1/2}$ by about $10\%$.

In Figures \ref{fig:vcircover} and \ref{fig:vcircunder} we show the classical dSphs in black and the ultra-faint dSphs in green. The dotted lines correspond to regions where the subhalos are not resolved, containing less than 200 particles. 

The increase in cross section from 1 to 5 \cmg does not visibly change the velocity profiles of the most massive subhalos outside 50 kpc. Within 50 kpc, we see a slight decrease in $V_{\rm{circ}}$ and $r_{\rm{max}}$ as cross section increases. 

In both the S1 and S5 simulation, the low density classical dSphs can be hosted by the simulated subhalos. However, the high density of Draco cannot be explained, as was also found in previous studies \citep{Valli:2017ktb,Read:2018pft}. We also find that Leo II, Sculptor, and Ursa Minor cannot be hosted by our S1 or S5 subhalos.
Of the five ultra-faint dSphs that we show, two can be hosted by both the inner and outer SIDM subhalos, one can be accommodated by the outer subhalos only, and two cannot be allocated in any of the subhalos plotted. 
Overall, the SIDM subhalos in Figures \ref{fig:vcircover} and \ref{fig:vcircunder} cannot host all of the observed dSphs. 

The overall subhalo abundance may change with MW halo mass and we have accounted for such variance in the subhalos structural parameters by modelling circular velocity curves for subhalos with larger $V_{\rm{max}}$ and $r_{\rm{max}}$ values. We found that the $V_{\rm{circ}}$ values at 0.25 kpc were no larger than those we see in Figure \ref{fig:vcircover}. Thus, SIDM simulations with larger host halo masses will not necessarily create subhalos that can host the densest dSphs.

CDM simulations, both with and without a disk, predict the existence of subhalos that are more dense than the brightest dSphs that we have observed (see e.g. Figure 3 of \citet{Robles:2019mfq, BoylanKolchin:2011dk, Garrison-Kimmel:2014vqa}), which motivated the TBTF problem. SIDM models can lead to the opposite problem: not predicting dense enough subhalos to explain the observed bright dSphs \citep{Robles:2019mfq, Read:2018fxs}. 
However, SIDM subhalos that undergo gravothermal core collapse are able to explain the densest dSphs \citep{Kaplinghat:2019svz, Kahlhoefer:2019oyt, Nishikawa:2019lsc}. 
In Figures \ref{fig:vcircover} and \ref{fig:vcircunder}, we do not see any subhalos that can explain the densest dSphs, which is the first indication that subhalos in the S1 and S5 simulations are not experiencing core collapse. We discuss this further in Section \ref{sec:corecollapse}.

We now take a closer look at the comparison between CDM, S1, and S5. Figure \ref{fig:VmaxRmax} shows the maximum circular velocity, $V_{\rm{max}}$, and radius of maximum circular velocity, $r_{\rm{max}}$, of all of the subhalos in Figures \ref{fig:vcircover} and \ref{fig:vcircunder}, color coded by their pericenters. $V_{\rm{max}}$ and $r_{\rm{max}}$ are calculated from the $z$=0 particle data and pericenter is the subhalo's closest approach to the center of the host halo over all time, calculated using energy and angular momentum conservation, as detailed in Section \ref{sec:subperi}. We observe a trend that subhalos with smaller pericenters have smaller $r_{\rm{max}}$ for a given $V_{\rm{max}}$. A similar trend is also mentioned by \citet{Robles_2021}, who characterize this trend by a power-law fit: $(r_{\rm{max}}/1\rm{kpc}) = A(V_{\rm{max}}/10\rm{kms}^{-1})^{1.5}$ where A depends on the pericenter. 

We find little difference in the $V_{\rm{max}}$ and $r_{\rm{max}}$ values of the S1 and S5 subhalos. To compare this to subhalos in CDM, we plot the power-law fit for all subhalos in the CDM+disk simulation in black in Figure \ref{fig:VmaxRmax}. We find that subhalos in CDM and SIDM simulations have similar $V_{\rm{max}}$ and $r_{\rm{max}}$ values.

\emph{Thus, the largest S5 subhalos can account for observed dSphs as well as S1 subhalos do. While subhalos in both simulations can account for the low density classical satellites, they have trouble matching the majority of the satellites because they don't scatter to sufficiently high densities. We see no significant difference in the $V_{\rm{max}}$ and $r_{\rm{max}}$ values of the subhalos in CDM, S1, and S5 simulations.
}

\section{Mass Loss}
\label{sec:massloss}

\begin{figure}
	\includegraphics[width=\columnwidth]{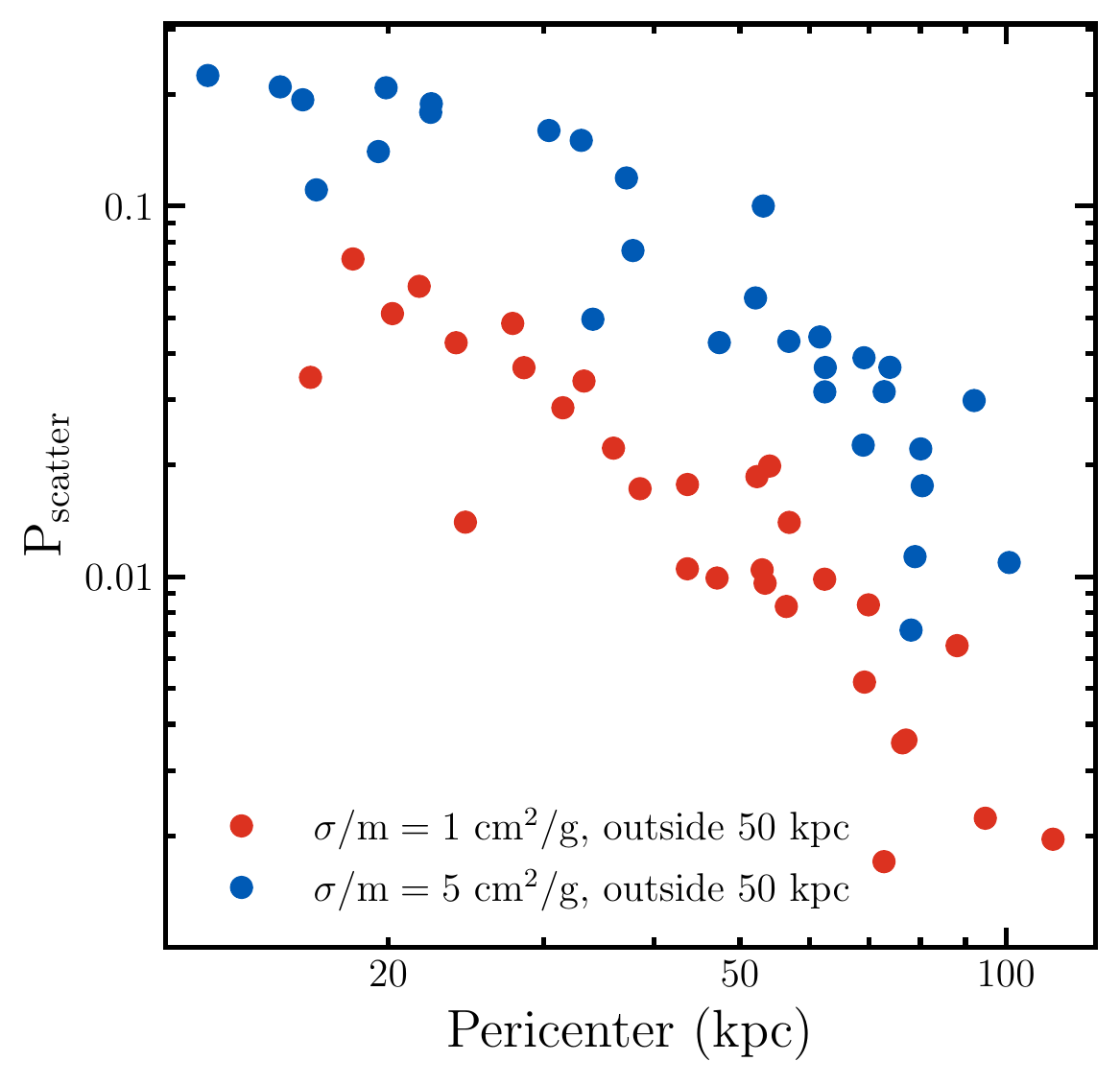}
    \caption{Probability that a single subhalo particle will scatter off a host halo particle in the subhalo's lifetime. We show this versus pericenter because subhalos with smaller pericenters traverse more dense regions in the host halo. We show the number of scatters for a particles in the 30 resolved subhalos currently outside 50 kpc with largest \Vpeak.}
    \label{fig:scatterrate}
\end{figure}

The simulations that we analyze assume a velocity-independent cross section. However, a more general model of self-interactions includes a velocity dependence such that scattering rate is inversely proportional to velocity~\citep{Feng:2009, Loeb:2011, Tulin:2013, Boddy:2014, Kaplinghat:2015aga, Tulin:2018}. Thus, scattering interactions that happen at high relative velocities occur more frequently in our simulation than in a model with velocity-dependence. The highest velocity particle collisions that occur in our simulations are between subhalo and host halo particles as the subhalo orbits the host halo at velocities of order 100 km/s. These scattering events effect the mass and trajectory of the subhalo \citep{Kummer_2017}.

First, we explore the measurable effects of the change in the subhalo's trajectory. After the host halo and subhalo particles interact, assuming the interacting particles have a large enough velocity, the host particles will escape the subhalo. The escaping particles decelerate in the subhalo's gravitational field, effectively transferring some of their momentum to the subhalo. Since the scattering particles are generally moving in the opposite direction as the subhalo, this momentum transfer decelerates the subhalo. When an orbiting subhalo decelerates, it moves radially inward towards the host halo. To measure this orbital change, we look at the pericenter of subhalos in our simulation. 

As shown in Figure \ref{fig:subperihist}, CDM has a slightly flatter distribution of pericenters, with fewer subhalos at the peak pericenter and more at the two ends. We see no measurable difference between the pericenters of subhalos in the S1 and S5 simulation, indicating that subhalo-host halo particle interactions have a small effect on the subhalos trajectory for cross sections of 1 and 5 \cmg. 

To quantify the effects on the mass of a subhalo due to a velocity independent cross section, we calculate the approximate mass lost by a subhalo due to scattering interactions between subhalo and host halo particles.

The interaction rate of a single particle within a subhalo is given by 
\begin{equation}
\Gamma = \frac{\sigma}{m} \langle \rho_{\rm{host}}(r) V_{\rm{rel}}(r) \rangle
\label{eq:intrate}
\end{equation}
where $\sigma$ is the cross section, and in our case $\sigma/m =$ 1 or 5 \cmg, $\rho_{\rm{host}}(r)$ is the background density of the host halo at the position of the subhalo and $V_{\rm{rel}}(r)$ is the relative velocity between the host halo and the subhalo. We calculate the average density times velocity as:
\begin{equation}
\langle \rho_{\rm{host}}(r) V_{\rm{rel}}(r) \rangle = \frac{\sum_{\rm{i}} m_{\rm{i}} \Delta V_{\rm{i}}}{\mathcal{V}}.
\label{eq:avedenvel}
\end{equation}
Here $\Delta V_{\rm{i}} = V_{\rm{i}} - V_{\rm subhalo}$ is the velocity of the particle with respect to the subhalo's velocity. We sum over all particles, i, within a sphere of radius $r = 1$ kpc that have velocity $V_{\rm i}> 3 V_{\rm{max, sub}}$. Both $V_{\rm subhalo}$ and $r_{\rm{vir, sub}}$, the virial radius of the subhalo, are determined by merger tree data. By excluding all particles that have a velocity less than three times the subhalo maximum circular velocity, we are cutting out the particles that are gravitationally bound to the subhalo. Assuming this cut removes \emph{all} subhalo particles, the density we calculate is that of the host halo at the location of the subhalo. $\mathcal{V}$ represents the volume of the sphere of interest, calculated as $\frac{4}{3} \pi r^3$, where $r$ is taken to be 1 kpc. To check our results, we also calculated the scattering probability using a volume of radius $r=5$ kpc and we find the same results.

The probability that a single subhalo particle will scatter off a host halo particle is given by integrating equation \ref{eq:intrate} over the subhalo's full orbit. For convenience, we perform the integral from the time a subhalo is accreted into the virial radius of the host until today.
\begin{equation}
    P_{\rm{scatter}} = 1 - \exp(- \int_{t_{\rm{acc}}}^{t_{\rm{today}}} \Gamma dt) \approx \int_{t_{\rm{acc}}}^{t_{\rm{today}}} \Gamma dt.
\label{eq:pscatter}
\end{equation}
This approximation can be made for small $\int \Gamma dt$.

Figure \ref{fig:scatterrate} shows the probability of scattering of a particle in a given subhalo from when it first enters the host halo's virial radius until today. We plot this as a function of the pericenter of that subhalo. Shown are the probability that a particle scatters in the 30 subhalos with largest \Vpeak that are located outside 50 kpc from the center of the host halo. This figure shows, as expected, that scattering rate is proportional to density and cross section. 

Figure \ref{fig:scatterrate} shows that our subhalos lose, on average, 2\% (9\%) of their mass in our S1 (S5) simulation.
A subhalo with close pericentric passage, of about 20 kpc, in the S1 simulation will lose at most 7\% of its DM particles. The larger cross section in the S5 simulation causes subhalos to lose more mass. A subhalo with close pericentric passage in the S5 simulation loses up to 23\% of its mass. While a larger mass loss does make subhalos more likely to undergo tidal stripping, Figure \ref{fig:subperihist} shows that pericenter alone does not lead to a significant difference in the number of subhalos that become fully tidally disrupted in the S1 versus the S5 simulations.

A similar analysis was conducted by \citet{Nadler:2020ulu} for subhalos in their dark matter only simulations. They conclude that with an isotropic cross section of 2 \cmg, subhalos with pericenters $\lesssim$ 70 kpc can lose about 10\% of their infall mass due to host halo-subhalo particle interactions. Comparatively, subhalos with pericenters $\approx $ 70 kpc in our S1 (S5) simulation lose about 0.8\% (4\%) of their mass. This difference in mass loss may relate to the fact that we simulate the gravitational effects of a baryonic disk and bulge, while \citet{Nadler:2020ulu} do not. The inclusion of the disk and bulge in MW simulations has been found to destroy subhalos with small pericentric passages.

It is notable that the three subhalos in these populations with smallest pericenters are all in the S5 simulation. To investigate these three subhalos further, we referred to their density profiles. These density profiles do not stand out from those of other subhalos, as is apparent from their $V_{\rm{max}}$ and $r_{\rm{max}}$ values. Starting with the subhalo with smallest pericenter, these subhalos have $V_{\rm{max}} = 8.75, 11.16, 15.03$ km/s, and $r_{\rm{max}} = 0.48, 2.14, 1.44$ kpc.

\emph{We conclude that mass loss due to subhalo-host halo interactions causes subhalos to lose, on average, 2\% (9\%) of their mass in our S1 (S5) simulation.}

\section{Core Collapse}
\label{sec:corecollapse}

\begin{figure*}
	\centering
	\includegraphics[width=\textwidth]{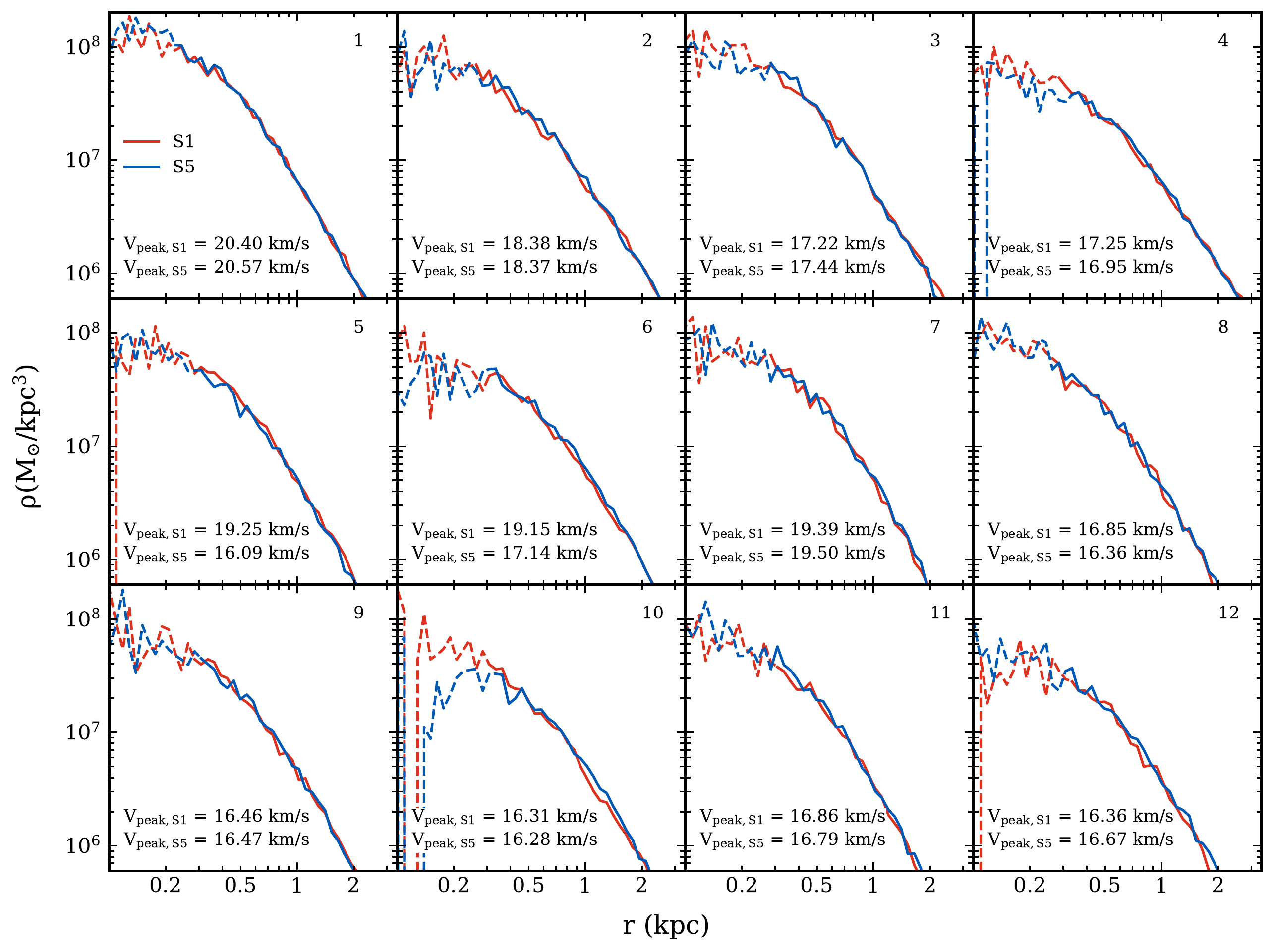}
    \caption{Density profiles for 12 pairs of matching subhalos from the S1 and S5 simulations. Pairs are found by maximizing equation \ref{eq:pairing} and only using pairs with $\xi > 0.3$. Dashed lines show unresolved regions (with less than 200 particles). Pairs are ordered from largest to smallest $V_{\rm{max}}$ of the S1 halo pair. \Vpeak values for both subhalos are given in the bottom left of each subplot and the values for  $V_{\rm{max}}$, and $r_{\rm{max}}$ are given in Table \ref{tab:pairinfo}.}
    \label{fig:denpairs}
\end{figure*}

\begin{table}
    \centering
    \begin{tabular}{c c c c c c c}
        Pair \#& \multicolumn{2}{c}{$V_{\rm{max}}$} & \multicolumn{2}{c}{$r_{\rm{max}}$} & \multicolumn{2}{c}{$r_{\rm{peri}}$}\\
         & \multicolumn{2}{c}{(\kms)}  & \multicolumn{2}{c}{(kpc)}& \multicolumn{2}{c}{(kpc)} \\
        \hline
          & S1 & S5 & S1 & S5 & S1 & S5 \\
         \hline
        1 & 18.41 & 18.59 & 1.02 & 0.94 & 48.40 & 47.84\\
        2 & 16.64 & 17.06 & 1.65 & 1.41 & 77.50 & 80.40\\
        3 & 16.60 & 16.49 & 0.94 & 0.94 & 88.04 & 92.05\\
        4 & 16.22 & 16.38 & 1.79 & 1.52 & 113.81 & 67.77\\
        5 & 15.74 & 15.31 & 1.02 & 1.02 & 56.06 & 55.00\\
        6 & 15.67 & 16.31 & 1.30 & 1.41 & 59.46 & 75.65\\
        7 & 15.63 & 15.74 & 1.02 & 1.11 & 52.27 & 53.14\\
        8 & 14.75 & 14.86 & 0.94 & 1.11 & 64.35 & 69.08\\
        9 & 14.62 & 14.74 & 1.11 & 1.02 & 69.88 & 73.90\\
        10 & 14.28 & 14.56 & 1.02 & 1.52 & 96.97 & 101.75\\
        11 & 13.55 & 13.88 & 0.94 & 0.94 & 38.94 & 39.88\\
        12 & 12.83 & 13.29 & 1.20 & 1.41 & 54.77 & 63.23\\
    \end{tabular}
    \caption{$V_{\rm{max}}$, $r_{\rm{max}}$, and pericenter values for the S1 and S5 subhalo pairs shown in Figure \ref{fig:denpairs}. The subhalos' pair number is given based on S1 $V_{\rm{max}}$ and is shown in the top right corner of Figure \ref{fig:denpairs}. Pairs are selected by maximizing $\xi$, a value that quantifies the number of particles that two subhalos share, and we only show pairs with $\xi > 0.3$.}
    \label{tab:pairinfo}
\end{table}

In a DM model with a small cross section ($\sigma/m = 1$ \cmg), DM particles undergo fewer collisions than in a model with a larger cross section ($\sigma/m = 5$ \cmg). Thus, subhalos in our S1 simulation undergo less mass loss due to host and subhalo particle interactions (as seen in Figure \ref{fig:scatterrate}). One might expect S1 subhalos to be more dense than, or at least as dense as a subhalo in the S5 simulation with similar initial conditions and orbit.

Another physical process called gravothermal core collapse \citep{Zel'dovich1966, lynden-bell_gravo-thermal_1968, larson_method_1970, Lightman1978, Shapiro1985} may also affect the densities of subhalos in our simulations. Through heat transfer to the colder, central regions of subhalos, core collapse increases the central densities of subhalos while decreasing the radius of the core \citep{hachisu_gravothermal_1978, lynden-bell_1980, Balberg:2002ue, balberg-SMBH-2002, Koda:2011}.
This is especially pertinent in DM models with large cross sections ($\geq$ 5 \cmg) at small velocities (order $V_{\rm{max}}$) \citep{Nishikawa:2019lsc, Turner:2021}. Core collapse is dependent on subhalo concentration, orbit, and DM self-interaction cross section \citep{Kahlhoefer:2019oyt, Sameie:2019zfo}. Furthermore, core collapse may cause the central densities of subhalos in models with large cross section to be larger than corresponding subhalos in models with small cross section (see \citet{Kahlhoefer:2019oyt} Figure 1).

In this section, we aim to determine whether subhalos in our S5 simulation have undergone core collapse. In Section \ref{sec:match} we match S1 subhalos to the S5 subhalos that have the most similar initial conditions. We find, however, that our resolution is not sufficient to make conclusions about core collapse. In Section \ref{sec:iso}, to explore subhalo densities within the resolution limit, we utilize an isothermal model. After confirming that the model correctly predicts subhalo densities in resolved regions, we use the model to extrapolate our density profiles to lower radii.

\subsection{Halo Matching}
\label{sec:match}
To determine if core collapse is playing a major role in our simulations, we look at direct comparisons of subhalos in the S1 simulation and the S5 simulation. To find these halo matches, we look at only the 30 subhalos within the outer region of the host halo (between 50 kpc and $R_{\rm{vir, host}}$) with largest \Vpeak, and compare subhalos in two ways. First, we look at the coordinate position of the subhalos at accretion, and evaluate matches assuming that corresponding subhalos would have similar accretion histories. 

Second, we look at the particles within each subhalo, which are flagged with a unique ID number. Since the S1 and S5 simulations were run from the same initial conditions, corresponding subhalos in both simulations should share many particles. We quantify the amount of shared particles using the equation:
\begin{equation}
    \xi = \frac{N_{\rm{shared}}^2}{N_1 N_2}
    \label{eq:pairing}
\end{equation}
where $N_1$ and $N_2$ are the number of particles in each subhalo, and $N_{\rm{shared}}$ is the number of particles that are shared between the two subhalos \citep{Knollmann_2009}. To find subhalo pairs, we maximize this quantity and only look at pairs with $\xi > 0.3$. All of the pairs that we find have $0.3 < \xi < 0.55$. 

Under these conditions, we found 12 pairs of subhalos. A more extensive search through all subhalos to find more pairs may be insightful in the future, but was not necessary for the analysis presented in this work.

With the subhalo pairs, we are able to compare how subhalos evolve in the S1 versus the S5 simulation. To compare the potential mass loss and core collapse in each model, we look at the $z$=0 density profiles. Figure \ref{fig:denpairs} shows the local density in spherical shells of each of the 12 pairs of subhalos. The S1 subhalos are in red and the S5 subhalos are in blue. The pairs are ordered from largest to smallest $V_{\rm{max}}$ of the S1 subhalo and the $V_{\rm{max}}$, $r_{\rm{max}}$ and $r_{\rm peri}$ are given in Table \ref{tab:pairinfo}. The solid line shows fully resolved areas of the density profile, while dashed lines indicate unresolved ($\leq$ 200 particles) radii. 

Figure \ref{fig:denpairs} shows that the density profiles of an S1 and S5 pair are very similar. All pairs have the same profile outside about 0.5 kpc, and only one pair, pair 10, differs significantly within 0.5 kpc. With the given resolution, we speculate that this subhalo pair had an orbit with few and far pericentric passages and has a low concentration such that neither subhalo underwent core collapse. Indeed, we confirmed that this subhalo pair has the largest pericenter of all 12 pairs.

From Figure \ref{fig:denpairs} we conclude that no subhalos in the S5 simulation that we have explored are undergoing core collapse to the extent that their densities increase to be larger than their S1 pair. We do see that some subhalos may have started to undergo core collapse more than others. For example subhalo pair 10, with the largest pericenter, doesn't seem to have undergone any core collapse, versus subhalo pair 11, with the smallest pericenter, where we see that the S5 subhalo density is at least as large at its S1 pair. However, we are limited by resolution at small radii.

\emph{The density profiles show no difference between S1 and S5 subhalos at the radii that are resolvable in our simulations ($r > 0.25$ kpc), indicating that our subhalos have not undergone core collapse.}

\subsection{Isothermal Model}
\label{sec:iso}
\begin{figure}
    \centering
    \includegraphics[width=\columnwidth]{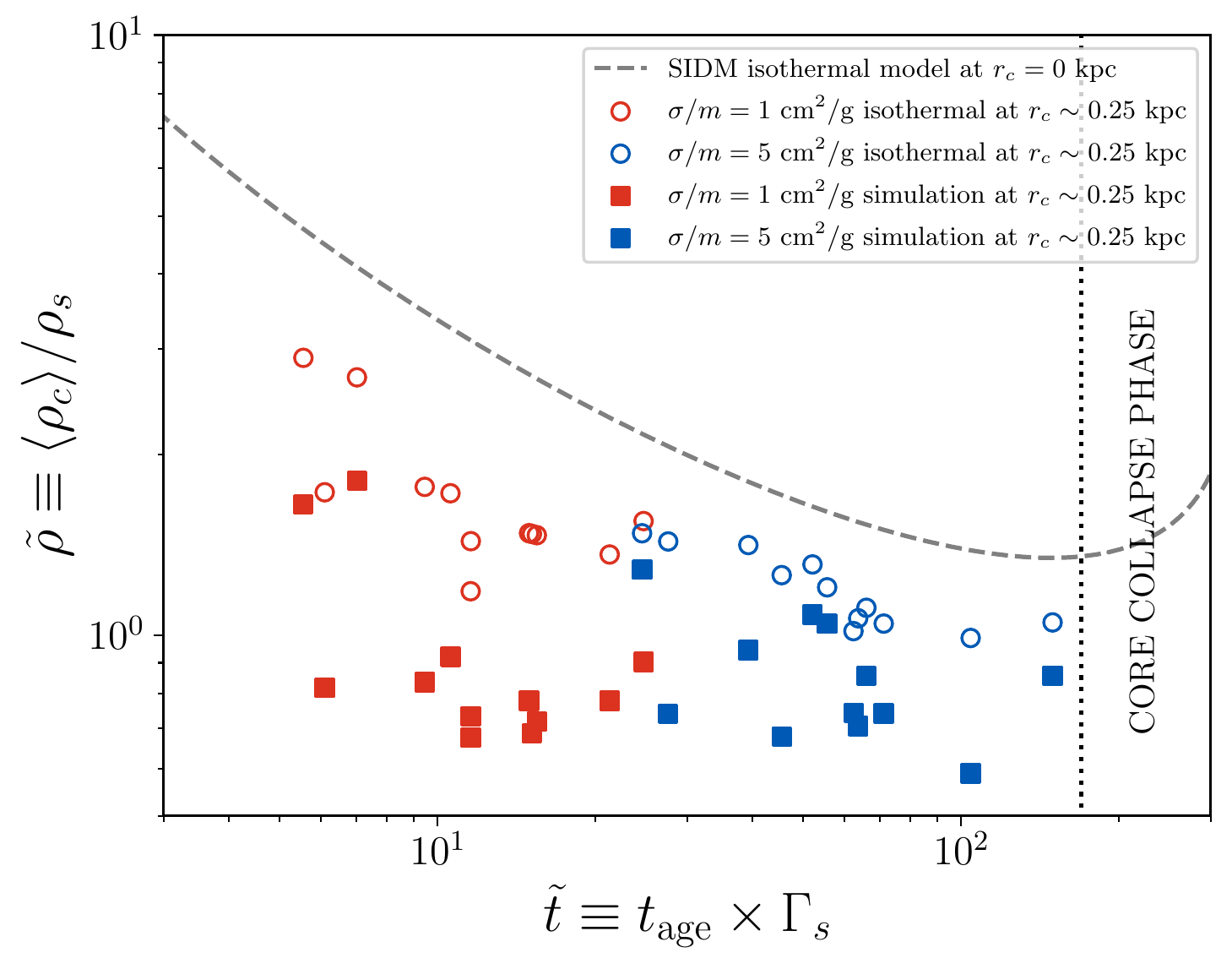}
    \caption{Average core density as a function of time. The squares show the densities and timescales of the subhalo pairs in the S1 (red) and S5 (blue) simulations. The circles show the predicted densities and timescales of the subhalo pairs, calculated from their $V_{\rm{max}}$ and $r_{\rm{max}}$ values. For both the squares and circles, we define the core radius, $r_{\rm{c}}$, as the innermost resolved radius, which is about 0.25 kpc. The dashed grey line shows the density at the center ($r$=0) of a subhalo predicted by the isothermal model. On the right side of the plot, we identify the times for which a subhalo has entered the core collapse phase, identified by an upturn in the dashed line. In this phase the core density of the subhalo increases.}
    \label{fig:corecollapse}
\end{figure}

\begin{figure*}
    \centering
    \includegraphics[width=\textwidth]{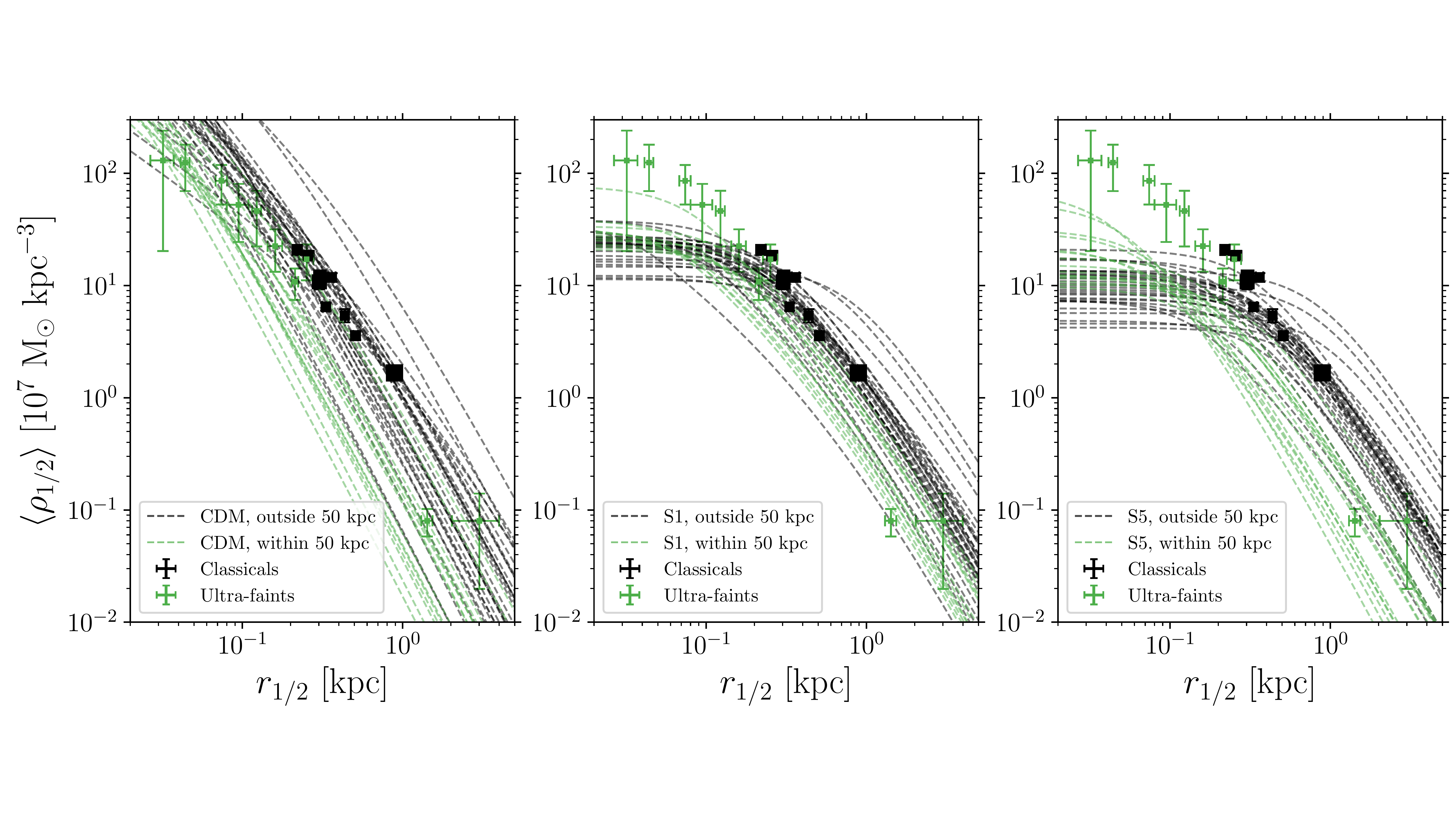}
    \caption{Mean density profiles of 30 resolved subhalos outside 50 kpc with largest \Vpeak (black) and all resolved subhalos within 50 kpc (green) in our S1 (middle) and S5 (right) simulation. From the CDM simulation (left), we show mean density profiles of 30 resolved subhalos outside 50 kpc with largest \Vpeak (black) and 15 resolved subhalos within 50 kpc with largest \Vpeak (green). Densities are calculated from an isothermal model with the $V_{\rm{max}}$ and $r_{\rm{max}}$ values from the simulations. We compare to the mean density and half-light radius of classical (black) and ultra-faint (green) dSphs. The size of the black squares is given by the luminosity of that dSph.
    }
    \label{fig:isothermalden}
\end{figure*}

In our SIDM simulations, we are limited by resolution and not able to make conclusions from the inner areas of subhalos, specifically, within about 0.25 kpc. Instead, we use an isothermal model to explore the central densities of subhalos in both SIDM models. We are particularly interested in two aspects of the central densities. First, from the central densities we determine whether our SIDM subhalos have entered the core collapse phase. We then compare the central density profiles to observed dSphs to address the TBTF problem highlighted in \citet{Kaplinghat:2019svz}.

The isothermal model that we use is described in detail in \citet{Kaplinghat:2013xca} (see also \citet{Kaplinghat:2015aga}). This model has been tested extensively by \citet{Robertson2020} who find it to be surprisingly accurate for SIDM models similar to ones we use in this work. We fit this isothermal model to our subhalos using their $V_{\rm{max}}$ and $r_{\rm{max}}$ values.

Figure \ref{fig:corecollapse} shows the average core density and how close our subhalo pairs are to the core collapse phase. The red squares show the density and timescales for the S1 subhalo pairs, and the blue squares show those for the S5 subhalos. Due to resolution constraints, the core density for our simulated subhalos is taken at the smallest radius at which both subhalos in a pair enclose at least 200 particles, which is about $r_{\rm{c}}$ = 0.25 kpc. The y-axis is the average core density divided by $\rho_{\rm{s}}$, which is taken to be $\rho_{\rm{s}} = 1.721 V_{\rm{max}}^2 G^{-1} R_{\rm{max}}^{-2}$ \citep{Kaplinghat:2019svz}. The x-axis is the age of the subhalo, taken to be 10 Gyr, multiplied by the scattering rate, which is $\Gamma_{\rm{s}} = a \frac{\sigma}{m} v_0 \rho_{\rm{s}}$ where $a = \sqrt{16/\pi}$, the velocity scale is $v_0 = \sqrt{G M_0/r_{\rm{s}}}$, and the mass scale is $M_0 = 4 \pi r_{\rm{s}}^3 \rho_{\rm{s}}$ \citep{Koda:2011,Nishikawa:2019lsc}. 

The circles show the average density at the core radius and timescale for the S1 and S5 subhalo pairs, predicted by the isothermal model. We calibrate the isothermal model using each subhalo's $V_{\rm{max}}$ and $r_{\rm{max}}$ values, and we plot the average density at that subhalo's $r_{\rm{c}}$.

Figure \ref{fig:corecollapse} shows that the isothermal model predicts average core densities that are slightly larger than the simulated subhalo core densities. The modeled and simulated subhalo densities agree better when only looking at the local density at $r_{\rm{c}}$ rather than the average density. The difference in average core density is thus due to differences in the density profiles in the central regions of the subhalos. This is likely due to the isothermal model deviating slightly from the subhalo density profile. An example of this can be seen in the right panel of Figure 4 in \citet{Robles:2019mfq}, which compares the density profiles of subhalos in the S1 simulation to the isothermal model. A more robust comparison between model and simulations in central the regions will be possible with higher resolution simulations.

On average, the model predicts larger core densities by up to a factor of about 2. This systematic error goes in the direction of making the densities predicted by the analytic model larger. We take this into account when we derive our conclusions regarding ultra-faint satellites.

 The dashed grey line shows the density evolution of the isothermal model at $r$ = 0 kpc. Because this shows the density at a different radius as the data points, they should not be directly compared. Rather, we show the density evolution of the isothermal model at $r$=0 to highlight the general trend of the density evolution; subhalos further along in their core collapse  timescales have lower average core densities. Additionally, we can use the time at which the isothermal model at $r$=0 begins to increase in density to estimate the time at which subhalos will enter the core collapse phase. This occurs when a subhalo's core density begins increasing. This is shown in Figure \ref{fig:corecollapse} as the region to the right of the dotted black line.

Figure \ref{fig:corecollapse} shows that no subhalos have entered the core collapse phase. This confirms that none of our subhalo pairs should show that the core density of the S5 subhalo is much larger than that of its S1 counterpart. 

To determine whether a difference in core radii could be observed in Figure \ref{fig:denpairs} we calculated the core radius, as defined by \citet{Kaplinghat:2013xca}, using the isothermal model. We find that the S5 subhalos have slightly larger core radii than their S1 pair. The predicted core radii for S1 subhalos lie between $0.15 - 0.23$ kpc, while the core radii of the S5 subhalos lie between $0.21 - 0.33$ kpc. Since the model predicts about a factor of two larger core densities, it likely also predicts smaller core radii.
A small shift in core radii, even for radii up to $0.5$ kpc would not be easily visible in Figure \ref{fig:denpairs}.
It is expected that core radii depend weakly on cross section for these subhalos, as is shown by the weak dependence on cross section of core density shown in Figure \ref{fig:corecollapse}.

Figure \ref{fig:isothermalden} shows the average density profiles from the isothermal model based on $V_{\rm{max}}$ and $r_{\rm{max}}$ values of subhalos in the S1 and S5 simulations (middle and right panels). We additionally show the average density profiles of subhalos in the CDM simulation (left panel). We show the profiles for the 30 subhalos currently outside 50 kpc with largest \Vpeak that are resolved at $r_{\rm{max}}$ (black dashed) and all subhalos currently within 50 kpc that are resolved at $r_{\rm{max}}$ (green dashed). Since CDM contains many resolved subhalos within 50 kpc, we show only those with the 15 largest \Vpeak values. We compare these to the half-light radius and average density at half-light radius of the classical and ultra-faint dSphs that we have been using throughout this paper. 

In the left panel of Figure~\ref{fig:isothermalden}, we see that CDM subhalos can explain all of the classical and ultra-faint dSphs well. However, there are a few massive simulated halos that should have created galaxies that we do not observe. This illustrates a tension, known as the TBTF problem; these subhalos are too big to fail at forming galaxies, yet we do not observe such galaxies.

While S1 and S5 somewhat alleviate the TBTF problem associated with classical dSphs, they both have trouble reproducing the high densities seen in the smallest ultra-faint dSphs. The S1 subhalos in the middle panel of Figure \ref{fig:isothermalden} can explain the observed classical dSphs because the most massive subhalos have large enough cores to host classical dSphs with small half-light radii. However, there are not enough subhalos that are dense enough to host all of the ultra-faint dSphs with $r_{1/2} < 200$ pc. The S5 simulation is even more discrepant with the ultra-faint population: there are four ultra-faint dSphs that are denser than {\em any} of the simulated subhalos.Additionally, at least two of the classical dSphs that are in tension with the subhalo velocity profiles (Figure \ref{fig:vcircover}) are also denser than the simulated subhalos in the S5 simulation. This tension is exacerbated by the fact that the isothermal model over predicts the average density of simulated subhalos by a factor of about two. 

Some values for mean density of dSphs may be updated as their binary fractions are taken into account. Using multi-epoch observations and Bayesian analysis, the binary fraction has been taken into account when calculating the velocity dispersion of Sculptor \citep{Minor2013}, Fornax \citep{Minor2013}, Carina \citep{Minor2013}, Sextans \citep{Minor2013}, Leo II \citep{Spencer2017}, Bo{\"o}tes I \citep{Koposov:2011zi}, and Segue I \citep{Martinez:2009jh}, but it remains to be done for others. Accounting for binary systems may change velocity dispersion, especially of ultra-faint dSphs \citep{Minor2010, McConnachie2010}, which affects the mean density and relative error of some dSphs, possibly decreasing the tension in Figure \ref{fig:isothermalden}.

Neither of our SIDM models are sufficient to explain both the classical and ultra-faint dSphs. 
An SIDM model with cross section significantly greater than 5 \cmg may be able to explain both the classical and ultra-faint dSphs with dense cores if some of the subhalos undergo core collapse. In a model with a large cross section, we anticipate that some of the outer subhalos will form larger cores, \emph{and} many subhalos will core collapse. A sharp velocity dependence to reduce mass loss in halo-subhalo dark matter scattering would likely be necessary to allow core collapse to proceed~\citep{zeng2021corecollapse}. Core collapsed subhalos may have cores dense enough to be compatible with the smallest ultra-faint dSphs. It would be interesting to test these expectations by running a Milky Way-like N-body simulation with a velocity dependent cross section that is greater than 5 \cmg at the subhalo circular velocity scale.

\emph{We conclude that neither the S1 nor the S5 simulations formed subhalos that are compatible with all the observed dSphs. Using an analytic model as an upper bound for the subhalo central densities, we predict that with a cross section at $V \approx V_{\rm{max}}$ of a factor of few larger, many subhalos would be in the core collapse phase and have large enough central densities to be consistent with the densest dSphs. This would require a sharp velocity dependence such that the effective cross section at orbital velocities ($V\sim 200~\rm km/s$) is below $\sim 1\rm cm^2/g$ to reduce mass loss from  halo-subhalo dark matter scattering.
}

\section{Conclusions}
\label{sec:conclusions}

We find that increasing the DM self-interaction cross section does not significantly change the density profile of a Milky Way-size host in N-body simulations that include a centrally-embedded baryonic potential that mimics the observed Milky Way (MW) galaxy today. While the central density (at regions within 10 kpc) does vary between models, the regions in which surviving subhalos orbit are very similar. The difference in central density does affect the circular velocity profile of the host halo such that the maximum circular velocity decreases from about 200 \kms in the CDM and low cross section (S1) simulations to about 150 \kms in the larger cross section simulation (S5). The host halos in both the S1 and S5 simulations are thermalized, while the host halo in the CDM simulation is not.

The number of subhalos produced in the SIDM simulations agree with observations, under the condition that subhalos in the S1 and S5 simulations with $V_{\rm{peak}} \gtrsim 7$ \kms are able to form stars. However, if the halos that can form stars is limited to those with \Vpeak > 20 km/s, as suggested by galaxy formation simulations, CDM DM-only, CDM with a baryonic disk, S1 with a baryonic disk, and S5 with a baryonic disk simulations are all in tension with the number of observed satellites.

To differentiate between DM models, we look at the subhalo pericenters, circular velocity profiles, and density profiles. With only minor differences in the subhalo pericenter distributions in the S1 and S5 simulations, we do not recommend this observable quantity be used to differentiate between the two models. In both SIDM models, we see a clear peak in pericenters between 20-40 kpc, while CDM has a slightly flatter distribution. 
All three DM models agree with observational data within Poisson errors, except within 20 kpc, where observed MW satellite galaxies outnumber simulated subhalos, and at about 60 kpc where simulated subhalos outnumber observed satellite galaxies. The mismatch beyond 60 kpc is likely affected by observational incompleteness. A more robust conclusion also depends on MW and LMC models.

When we look at the subhalo circular velocity profiles, we again see that subhalos in the S1 simulation are similar to those in the S5 simulation. Both simulations have subhalos that can explain some of the classical and a few ultra-faint dwarf-spheroidal satellite galaxies (dSph), however neither model can explain all observed dSph. When we look specifically at the maximum circular velocity ($V_{\rm{max}}$) and radius of maximum circular velocity ($r_{\rm{max}}$), CDM, S1, and S5 subhalos all lie in the same region.

Before looking at subhalo density profiles, we explore how much mass a subhalo loses due to host halo-subhalo particle interactions (dark ram-pressure stripping) that could lead to an observable difference in density profiles in S1 and S5 subhalos. We calculate the mass lost by a subhalo by integrating the scattering rate of a single particle over the subhalos trajectory after it enters the virial radius of the host (see equations \ref{eq:intrate}-\ref{eq:pscatter}). Because scattering probability is proportional to cross section, subhalos in our S5 simulation lose more mass through subhalo-host halo particle interactions than subhalos in our S1 simulation. While we do find that subhalos with the smallest pericentric passages in our S5 simulation lose about 23\% of their mass, the average subhalo in our S1 and S5 simulations loses 2\% and 9\% of their mass, respectively.

To observe the effects of mass loss, we compare the density profiles of corresponding subhalos in the S1 and S5 simulation. Our resolution due to DM particle mass limits the precision of these density profiles within a radius of about 250 pc. Outside this radius, there is no significant difference between S1 and S5 halos. Thus we conclude that although subhalos in our S5 simulation lose more mass than those in our S1 simulation due to host halo-subhalo interactions, this does not create observable differences in the pericenter distributions or in the outer regions of subhalo density profiles.

To probe the inner regions of these subhalos, we fit an isothermal model to the $V_{\rm{max}}$ and $r_{\rm{max}}$ values of the subhalos in our S1 and S5 simulations. We first confirm that the isothermal model predicts core density values (note that we define the core as the smallest resolved radius) that agree with the subhalos in both the S1 and S5 simulation up to a factor of two. Since the model consistently over predicts the subhalo central densities, we use the model as an upper bound for the subhalo central densities. By calculating the collapse timescale from each subhalo's $V_{\rm{max}}$ and $r_{\rm{max}}$ values, we also find that none of our subhalos are in the core collapse phase.

The key result, shown in Figure \ref{fig:isothermalden}, is that the inner density profiles of the modeled S1 and S5 subhalos are not dense enough to host observed ultra-faint dSphs with half-light radii smaller than $\sim 0.2$ kpc nor the most dense classical dSphs. It is possible that a cross section between 0 and 1 \cmg at the relevant subhalo velocity scale may predict dense enough ultra-faint dSphs while avoiding the TBTF problem. We cannot test this possibility here. Another avenue with more interesting phenomenological implications is an SIDM model with a cross section much larger than 5 \cmg at the subhalo circular velocity scale. Larger cross sections will cause core collapse to occur faster and would lead to some subhalos having high enough central densities consistent with the inferred densities in ultra-faint dSphs \citep{Turner:2021}. We note that dSph velocity dispersion data depends on the binary fraction within each satellite, which has not been accounted for in a few of the dSphs that we use in this work. Accounting for binary systems may change velocity dispersions, especially of ultra-faint dSphs \citep{McConnachie2010}, which affects the mean density of some dSphs, possibly decreasing the tension in Figure \ref{fig:isothermalden}.

Based on our results analyzing N-body simulations with a disk potential, SIDM models with cross sections between 1 and 5 \cmg are not consistent with the range of densities inferred for MW satellites. While subhalos that have undergone core collapse would be able to explain the densest satellites, we find that $\sigma/m = 5$ \cmg is insufficient to induce core collapse in subhalos within the age of the universe. To match the luminosity function of the satellites, we find that galaxies should form in halos with peak circular velocity as low as $7$ \kms for a cross section of $5~ \rm cm^2g^{-1}$, which is slightly lower than the threshold with no self-interactions. Our results, along with other analyses in the literature, strongly suggest that viable SIDM models should have a sharp velocity-dependence with cross section values much larger than 5 \cmg at the velocities relevant for MW dSphs falling to values much less than $1$ \cmg at velocities in galaxy clusters.

\section*{Acknowledgements}
We thank Mariangela Lisanti, Fangzhou Jiang, Oren Slone, and Ethan Nadler for useful discussions and helpful comments. VHR acknowledges support by the Yale Center for Astronomy and Astrophysics Prize postdoctoral fellowship. MV acknowledges support by U.S.~NSF Grant No.~PHY-1915005 and by the Simons Foundation under the Simons Bridge Fellowship, award number~815892. MK is supported by the NSF grant PHY-1915005. The simulations were run on the Texas Advanced Computing Center (TACC; http://www.tacc.utexas.edu).

\section*{Data Availability}

The data for simulated halo catalogs and merger trees in this article are publicly available in the article \citet{Kelley:2018pdy} for CDM. For SIDM the data are available on reasonable request to the corresponding author. Projected half-light radii data for observed dwarf spheroidal galaxies are taken from \citet{MNRAS:Munoz2018} and we deproject the radii by the conversion factor described in \citet{Wolf:2009tu}. For ultra-faint dwarfs we use the published data from \citet{MNRAS:Simon2019}. The pericenter distances used in this article were taken from published values in \cite{Patel_2020} and \citet{Fritz:2018aap}.



\bibliographystyle{mnras}
\bibliography{main} 




\appendix

\section{Time Dependent Host Halo Profile}
\label{sec:apphostprof}

While the density profile of the host halo does evolve with time, we use the NFW profiles at early ($z$=3) and late ($z$=0) times to show that the change is negligible in regions of highest interest. The subhalos that we examine throughout this paper, as well as most other subhalos in the simulations, do not ever come within 10 kpc of the center of the host halo. Thus, the region of interest is outside 10 kpc.

\begin{figure}
	\includegraphics[width=\columnwidth]{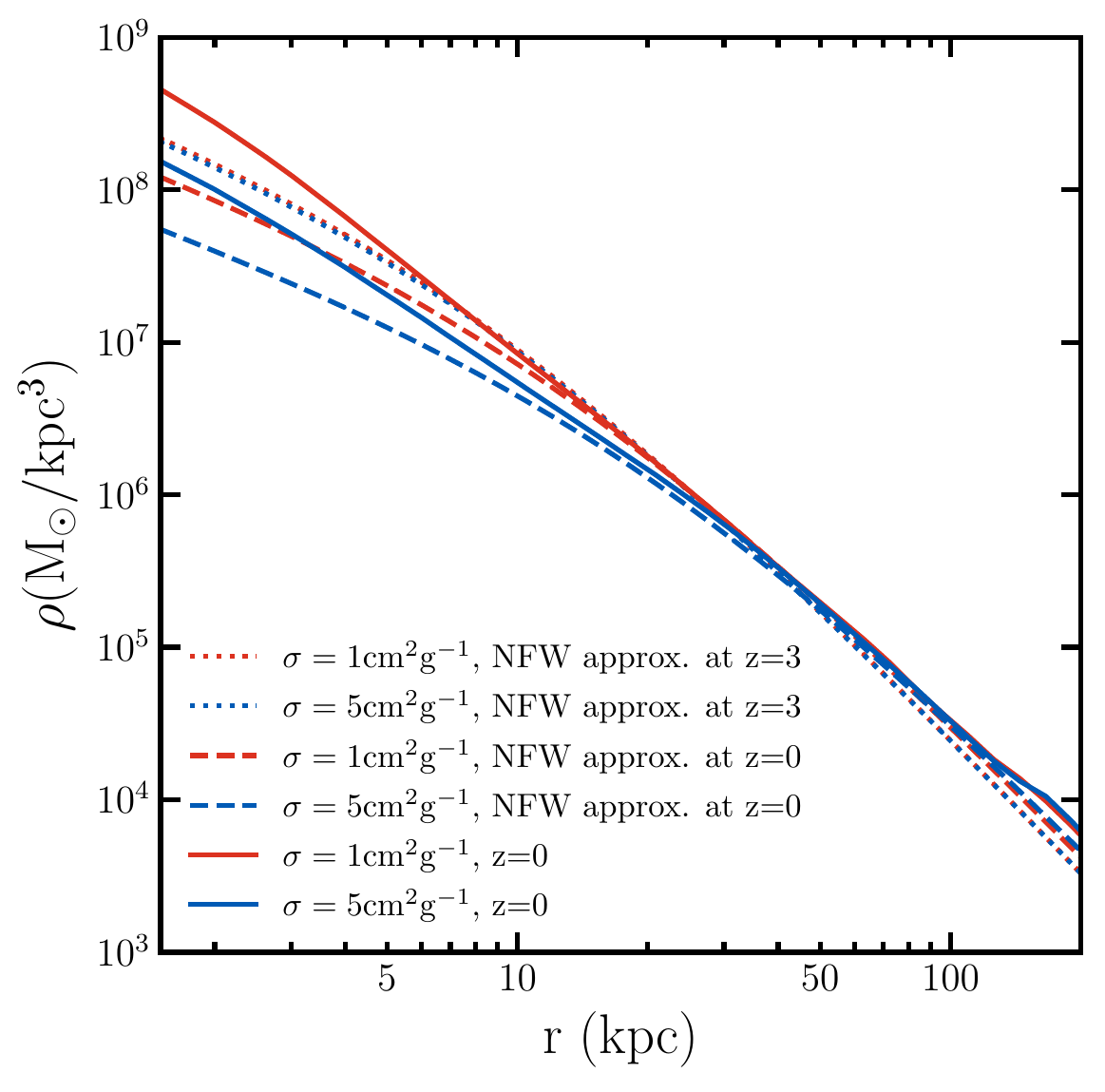}
    \caption{Host halo dark matter density. The density is calculated using the particle data, as described in Section \ref{sec:hostprof} (solid) and by calculating the approximate NFW profile (dashed). NFW profiles are calculated at $z$=0 (dashed) and $z$=3 (dotted) for $\sigma/m = 1$ cm$^2$g$^{-1}$ (red), and $\sigma/m = 5$ cm$^2$g$^{-1}$ (blue).}
    \label{fig:hostdenNFW}
\end{figure}

Figure \ref{fig:hostdenNFW} shows the density profiles of the host halo in our S1 (red) and S5 (blue) simulations. Particle data is used to calculate these profiles at $z$=0 (solid), and parameters from the merger trees are used to estimate NFW density profiles at $z$=0 (dashed) and $z$=3 (dotted).

Comparing the density profiles at different redshifts and at 10 kpc, the NFW approximation of the S5 halo has become more dense by about a factor of 1/2, while the S1 halo has remained at the same density. Outside 10 kpc, this difference in the S5 halo decreases to a negligible amount. Thus, we conclude that the S1 and S5 host halo density profiles throughout all time can be reasonably approximated by their particle density profiles at $z$=0.

\section{Pericenter Calculation}
\label{app:pericalc}

To calculate simulated subhalo pericenters with higher resolution than the snapshots captured by the simulation, we use a simple model that assumes conservation of energy and angular momentum between the two snapshots that straddle the pericentric passage. This refined interpolation only changes pericenter values by an average of 3\%, which does not affect our conclusions.

The angular momentum is $L = r V_{\rm tan}$, where $V_{\rm tan}$ is the tangential velocity of the subhalo with respect to the host halo and $r$ the relative distance. We approximate the gravitational potential as $A + Br$ and approximate $B=(\Delta V_{\rm tot}^2)/(2\Delta r)$
assuming energy ($=A+Br+V_{\rm tot}^2/2$) conservation. Pericentric passage occurs when the radial velocity is zero, implying that $V_{\rm tot}^2 = V_{\rm tan}^2 = L^2/r^2$. Rearranging this equation, we get:
\begin{equation}
    B r_{\rm peri}^3 - (E-A) r_{\rm peri}^2 + \frac{L^2}{2} = 0
\end{equation}
We use the smallest positive root of this equation as the pericenter.

There are a few cases where our assumptions do not hold. If the subhalo has not made a pericentric passage, we use the distance of closest approach captured by the simulation, which is either in the first or last snapshot. If the subhalo's angular momentum changes by more than 50\%, we use the distance of closest approach captured by the simulation. This may be the case if dynamical friction is changing the subhalo's orbit. If the gravitational potential is clearly non-spherical, resulting in a negative gradient of the gravitational potential, we again use the distance of closest approach captured by the simulation.

\section{Subhalo Velocity Distribution}
\label{sec:subveldisp}

\begin{figure}
	\includegraphics[width=\columnwidth]{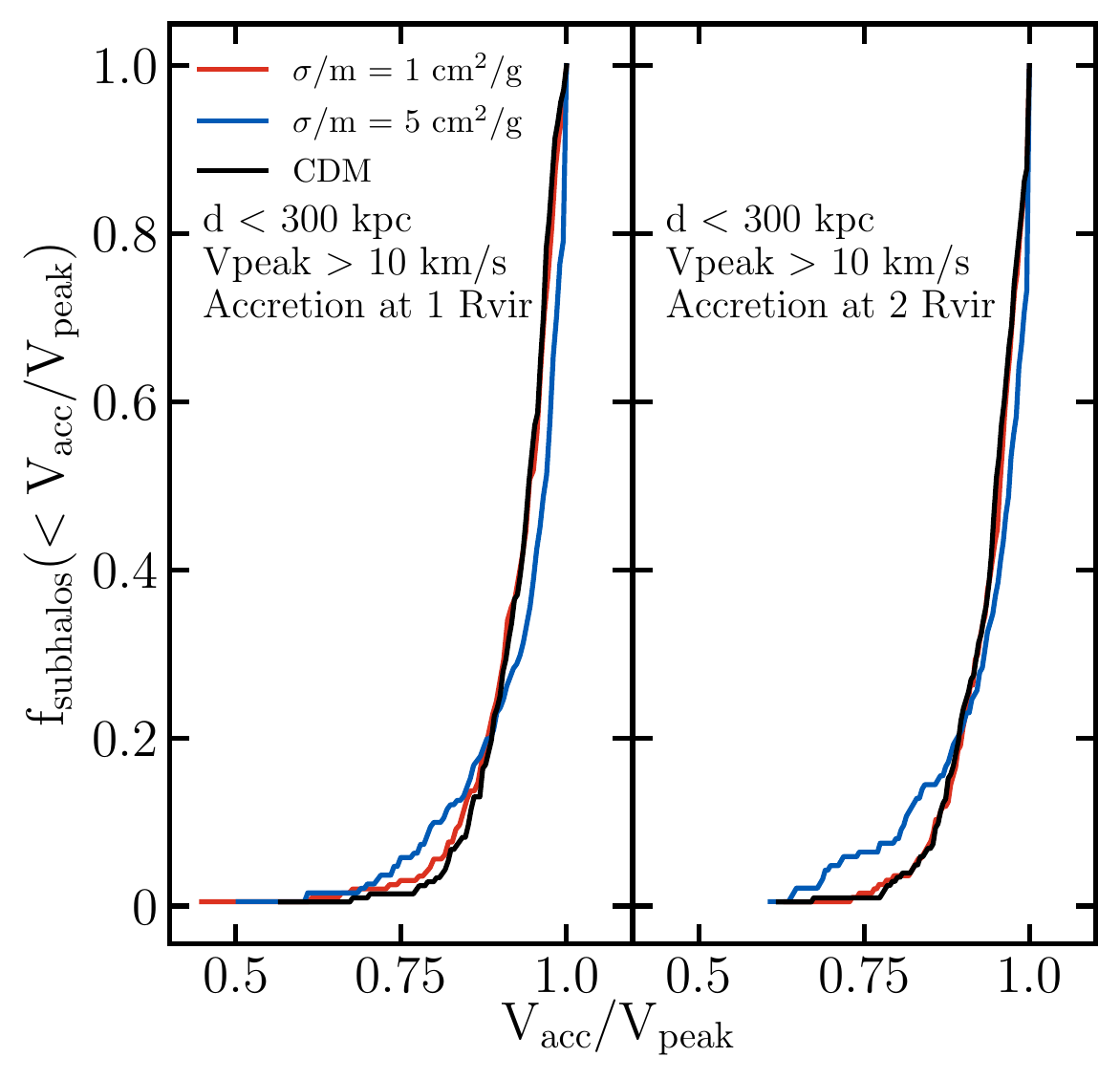}
    \caption{
    Normalized cumulative fraction of the maximum circular velocity at accretion divided by the largest maximum circular velocity of the subhalo before accretion. $f_{\rm{subhalos}} (< V_{\rm{acc}}/V_{\rm{peak}}) = N_{\rm{subhalos}}(< V_{\rm{acc}}/V_{\rm{peak}}) / N_{\rm total}$. The left panel shows the histogram when accretion is taken to be at 1 $R_{\rm{vir}}$ and the right panel shows the histogram when accretion is taken to be at 2 $R_{\rm{vir}}$.
    }
    \label{fig:subvaccvpeakhist}
\end{figure}

To get a better idea of how the host halo affects orbiting subhalos, we look at subhalo characteristics before and at accretion. We use the maximum circular velocity at accretion, $V_{\rm{acc}}$ and the peak circular velocity, \Vpeak as indicators of mass at these times. 

We define accretion as the earliest time at which a subhalo is within the virial radius of the host halo. If a subhalo never comes within $R_{\rm vir}$, accretion is taken to be at $z$=0. This is a valid approximation because we only look at subhalos that are currently within 300 kpc of the host halo, which has a virial radius of about 270 kpc. 

\Vpeak is often defined as the largest maximum circular velocity over all time and corresponds closely with subhalo mass. Self interactions as well as tidal stripping cause halos to lose mass while they are inside the virial radius of the host halo, so \Vpeak occurs before the subhalo has been accreted (except for some cases with major mergers). To correct for nonphysical jumps in the $V_{\rm{max}}$ data, we define \Vpeak as the largest $V_{\rm{max}}$ before accretion. 

The ratio of $V_{\rm{acc}}$ and \Vpeak describes the change in mass that subhalos experience before entering the host halo. 

The left panel of Figure \ref{fig:subvaccvpeakhist} shows a cumulative, normalized histogram of the ratio of $V_{\rm{acc}}$ and \Vpeak for subhalos with \Vpeak > 10 \kms. The tail at the low end of $V_{\rm{acc}}/V_{\rm{peak}}$ means that some subhalos in both SIDM simulations are losing as much as 50\% of their maximum mass before falling into the host halo. S5 has a larger fraction of subhalos that lie in the region $0.6 \leq V_{\rm{acc}}/V_{\rm{peak}} \leq 0.8$, but there is no noticeable difference between S1 and S5 below 0.6. 

Similarly, \citet{Nadler:2020ulu} observed a trend in the fraction of subhalos at the low end of the $V_{\rm{acc}}/V_{\rm{peak}}$ tail that is dependent on self-interaction velocity scale: larger velocity scales lead to a more prominent tail.

At first glance, it seems this trend may be explained by mass loss due to self interactions in the dense region outside the virial radius out to the splashback radius \citep{Diemer_2014}. For typical MW mass halos, the splashback radius is located at about 1.5 times the virial radius \citep{Adhikari_2014, More:2015}. To check this theory, the right panel of Figure \ref{fig:subvaccvpeakhist} shows a cumulative, normalized histogram of the ratio of $V_{\rm{acc}}$ and \Vpeak, with accretion defined to be at 2 $R_{\rm{vir}}$. While the very low end tail $V_{\rm{acc}}/V_{\rm{peak}}$ < 0.6 disappears in this case, we still see a larger fraction of S5 subhalos in the region $0.6 \leq V_{\rm{acc}}/V_{\rm{peak}} \leq 0.8$. We see similar results when looking at subhalos with \Vpeak > 4.5 \kms and \Vpeak > 15 \kms.

This is in contrast to what \citet{Nadler:2020ulu} report. They find  the difference between SIDM and CDM to decrease when defining accretion at larger radii, while we find that the difference is slightly amplified. 

While we do observe that subhalos in the S5 simulation lose more mass than those in the S1 and CDM simulations before accretion, the mass loss we observe is not due to self interactions between host halo and subhalo particles at the infall velocity scale. We are unclear why we find different results than \citet{Nadler:2020ulu}, which motivates further discussion. The different results may be due to simulation implementation, a difference in host halo mass, or a difference in cross section (we use a spherically averaged cross section while they use a differential cross section).

\section{Subhalo Density Profiles}
\label{sec:appdenprofs}
\begin{figure*}
	\includegraphics[width=\textwidth]{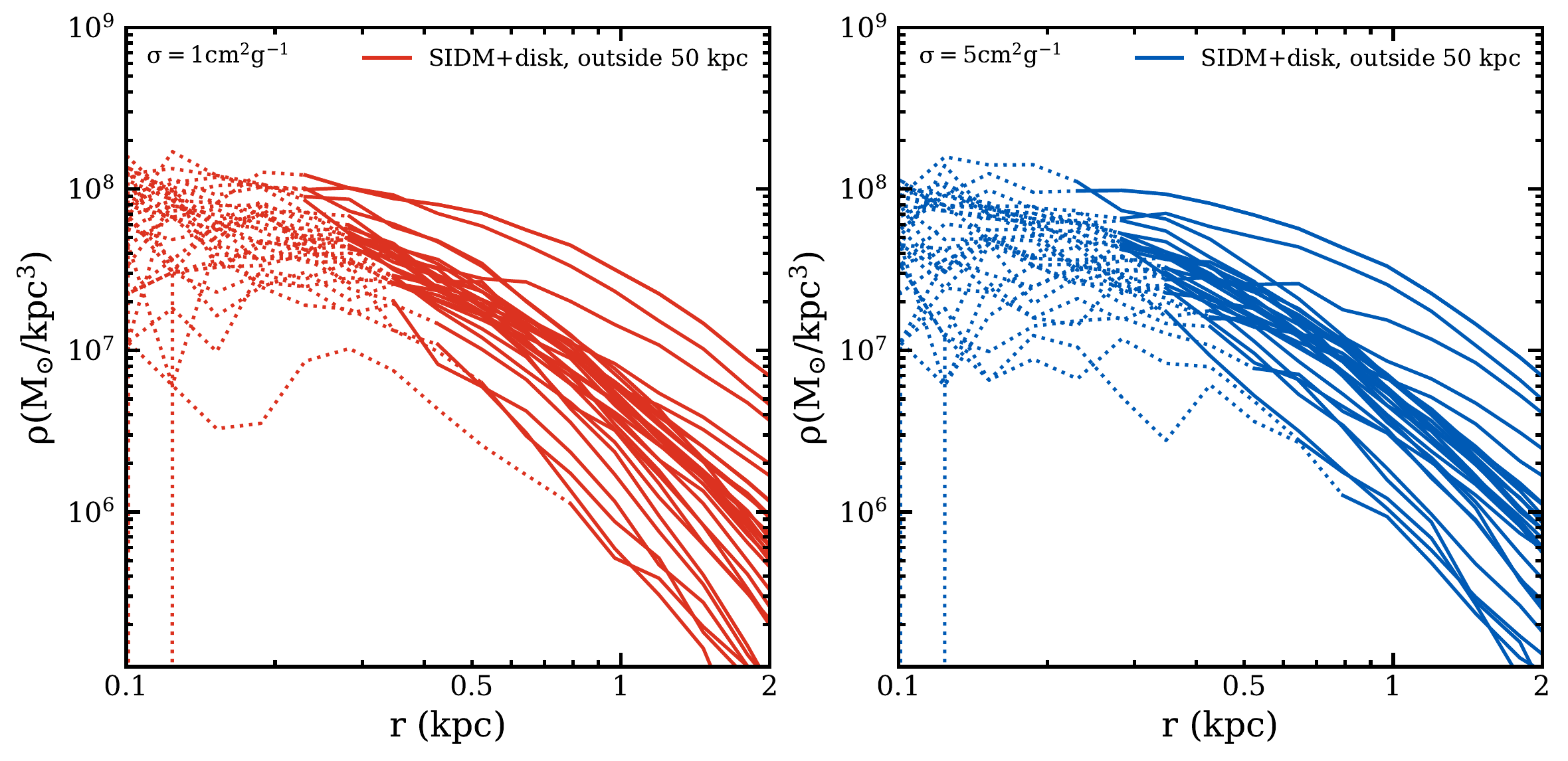}
    \caption{
    Density profiles of the 30 subhalos that are currently between 50 kpc and $R_{\rm{vir, host}}$ with largest $V_{\rm peak}$. Density is calculated as mass per volume in 20 logarithmically spaced bins. Dotted lines indicate radii that enclose less than 200 particles.}
    \label{fig:denover}
\end{figure*}

\begin{figure*}
	\includegraphics[width=\textwidth]{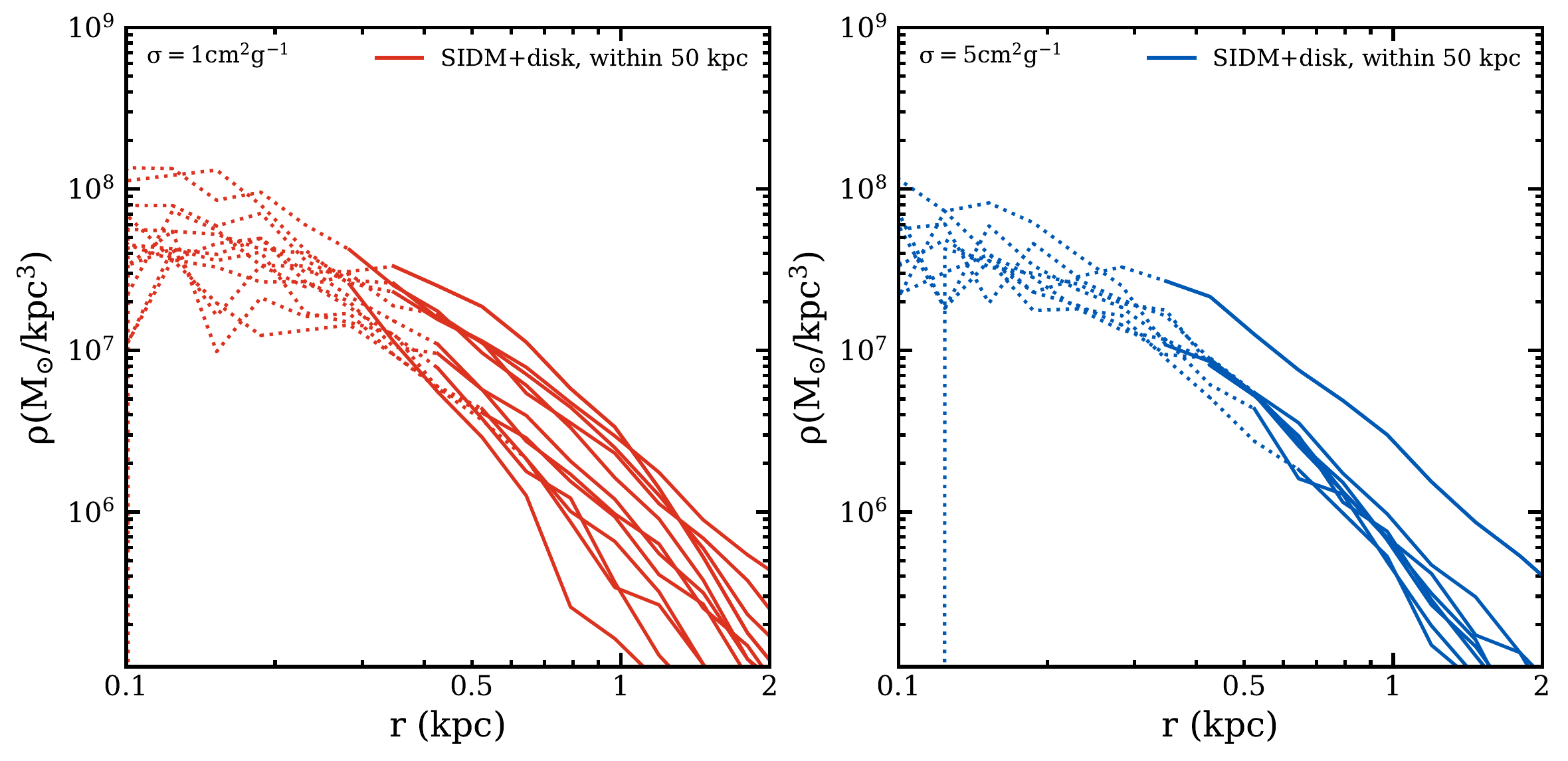}
    \caption{Density profiles of all subhalos that are currently within 50 kpc that are resolved at $r_{\rm{max}}$. Density is calculated as mass per volume in 20 logarithmically spaced bins. Dotted lines indicate radii that enclose less than 200 particles.}
    \label{fig:denunder}
\end{figure*}

Figure \ref{fig:denover} shows the density profiles of the 30 subhalos that are currently between 50 kpc and $R_{\rm{vir, host}}$ with largest \Vpeak that are resolved at $r_{\rm{max}}$. Figure \ref{fig:denunder} shows the density profiles of all subhalos currently within 50 kpc that are resolved at $r_{\rm{max}}$. Density is calculated as mass per volume within 20 logarithmically spaced bins. The dotted lines indicate radii within which there are less than 200 subhalo particles. 

We see no difference in the density profiles of S1 and S5 subhalos in the outer region (>50kpc). This conclusion is confirmed by Figure \ref{fig:denpairs}. For subhalos closer to the host (<50kpc), we again see no difference in their density profiles, but we do see that S5 halos have slightly less mass, resulting in larger unresolved regions. 

Although we are limited by resolution, we are able to suggest that outer subhalos (Figure \ref{fig:denover}) have cores that extend to larger radii than inner subhalos (Figure \ref{fig:denunder}). Many of the outer subhalos have cores that extend to about 0.5 kpc while the inner subhalos have cores that extend only to about 0.15 kpc. This observation is further illuminated in Figure \ref{fig:isothermalden} where we are able to use an isothermal model to extend the density profiles to smaller radii.

\section{Non Spherical symmetry in dwarf galaxy half-light radii}
\label{app:nosph}

In this section we show the S1 and S5 subhalo $V_{\rm{circ}}$ profiles compared to observed dSphs that have not been sphericalized. While we do not assume spherical symmetry in observed dSphs, we do still assume that our simulated subhalos are spherically symmetric, so we are hesitant to draw strong conclusions from this comparison. We plot the same subhalo circular velocity profiles as in Figures \ref{fig:vcircover} and \ref{fig:vcircunder} compared to the same dSphs shown in Figures \ref{fig:vcircover} and \ref{fig:vcircunder}. Table \ref{tab:dsph} lists the projected 2D half-light radius, $R_{\rm{e}}$, the deprojected 3D half-light radius, $r_{1/2}$, the ellipticity, and the sphericallized 3D $r_{1/2}$.

To compare the dSphs to the subhalo circular velocity profiles, we plot the dSph velocity at half-light radius, $V_{1/2}$ versus the 3 Dimensional deprojected half-light radius, $r_{1/2}$. We calculate $V_{1/2}$ using equation \ref{eq:v12} and the $\langle \sigma_{\rm{los}} \rangle ^2$ in \citet{MNRAS:Simon2019}. We use the $R_{\rm{e}}$ in \citet{MNRAS:Simon2019} and convert to $r_{1/2}$ using the conversion of 4/3 \citep{Wolf:2009tu}.

\begin{figure*}
    \centering
    \includegraphics[width=\textwidth]{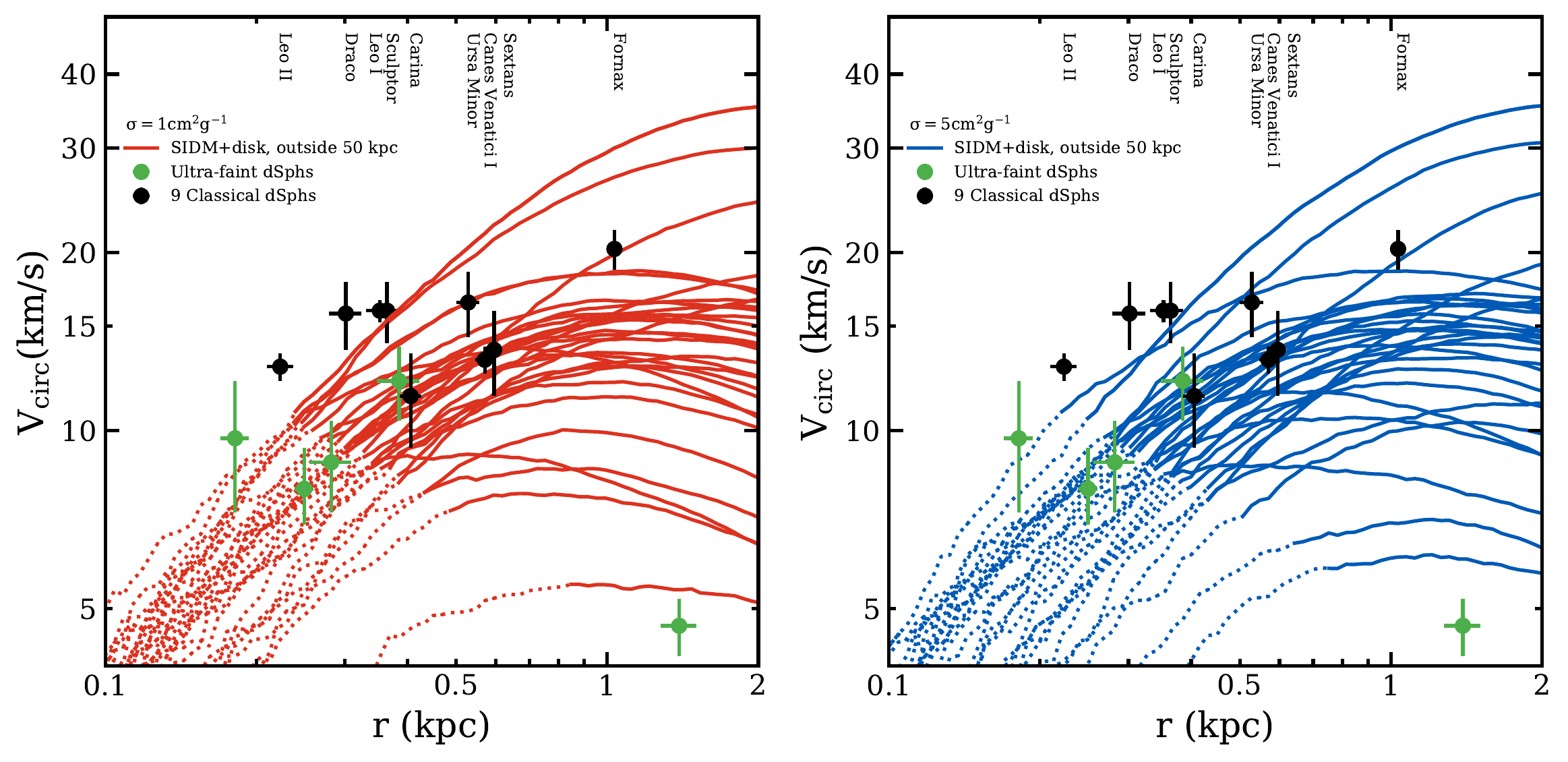}
    \caption{Circular velocity of the 30 subhalos that reside between 50 kpc and the host halo virial radius at $z$=0 with largest \Vpeak that are resolved at $r_{\rm{max}}$. We compare both S1 (left) and S5 (right) to the nine classical dSphs, which all reside outside 50 kpc from the MW center. We also compare to ultra-faint dSphs, some of which reside within 50 kpc and some which reside outside 50 kpc. We plot the dSph deprojected half-light radius, $r_{1/2}$ (given in Table \ref{tab:dsph}), and velocity at half-light radius, $V_{1/2}$ not assuming spherical symmetry.}
    \label{fig:vcircovernosph}
\end{figure*}

Figure \ref{fig:vcircovernosph} shows that both our S1 and S5 subhalos can house all but two of the classical dSphs and all but one of the ultra-faint dSphs plotted. Previous studies have also found that Draco is an outlier when compared with subhalos formed in SIDM simulations \citep{Valli:2017ktb, Read:2018pft}. We find that with the most recent data, Leo II is also in tension with our simulated subhalos. 

\begin{figure*}
    \centering
    \includegraphics[width=\textwidth]{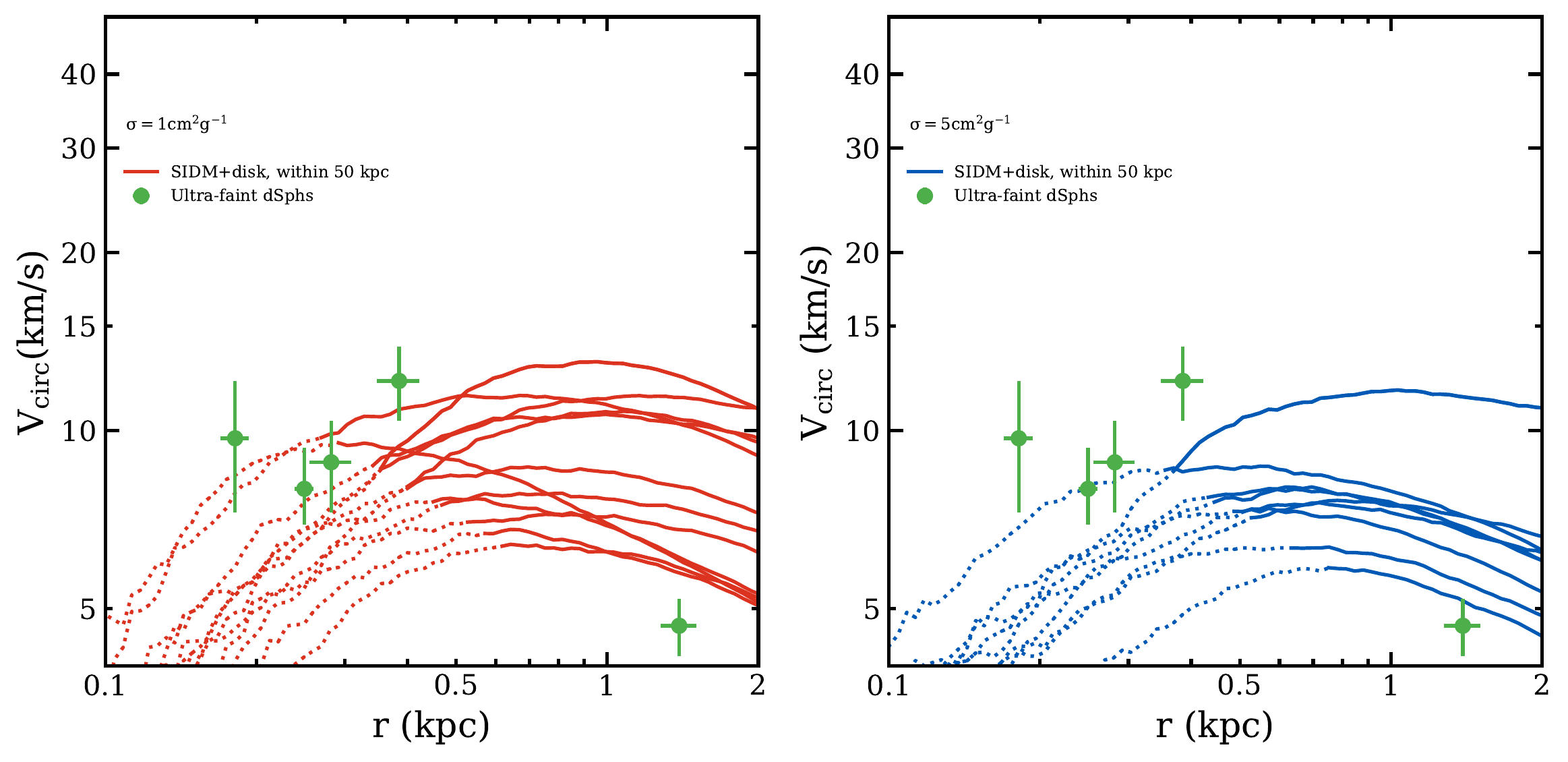}
    \caption{Circular velocity of all subhalos that reside within 50 kpc of the host halo center at $z$=0 that are resolved at $r_{\rm{max}}$. We have 12 resolved S1 subhalos and 8 resolved S5 subhalos. We compare to ultra-faint dSphs that reside within 50 kpc. We plot the dSph deprojected half-light radius, $r_{1/2}$ (given in Table \ref{tab:dsph}), and velocity at half-light radius, $V_{1/2}$ not assuming spherical symmetry.}
    \label{fig:vcircundernosph}
\end{figure*}

The left panel of figure \ref{fig:vcircundernosph} shows that the subhalos within 50 kpc in our S1 simulation can explain the observed ultra-faint dSphs well. However, the right panel of figure \ref{fig:vcircundernosph} shows that the subhalos within 50 kpc in our S5 simulation have trouble explaining all of the observed ultra-faint dSphs.

\begin{table}
    \centering
    \begin{tabular}{c c c c c}
         dSph & $R_{\rm{e}}$ & $r_{1/2}$ & $\epsilon$ & $r_{1/2}(1 - \epsilon)^{1/2}$\\
         & (pc) & (pc) & & (pc)\\
         \hline
         Sculptor & 279 & 372.0 & 0.33 & 304.50\\
         Fornax & 792 & 1056.0 & 0.29 & 889.80\\
         Carina & 311 & 414.7 & 0.36 & 331.73\\
         Leo I & 270 & 360.0 & 0.3 & 301.20\\
         Sextans & 456 & 608.0 & 0.3 & 508.69\\
         Leo II & 171 & 228.0 & 0.07 & 219.88\\
         CVn I & 437 & 582.7 & 0.44 & 436.03\\
         UMi & 405 & 540.0 & 0.55 & 362.24\\
         Draco & 231 & 308.0 & 0.29 & 259.53\\
         \hline
         UMa II & 139 & 185.3 & 0.56 & 122.94\\
         UMa I & 295 & 393.3 & 0.59 & 251.86\\
         Bo{\"o}tes I & 191 & 254.7 & 0.3 & 213.07\\
         Hercules & 216 & 288.0 & 0.69 & 160.35\\
         Cb I & 69 & 92.0 & 0.37 & 73.02 \\
         CVn II & 71 & 94.7 & 0 & 94.67\\
         Cra II & 1066 & 1421.3 & 0 & 1421.33\\
         Ret II & 51 & 68.0 & 0.58 & 44.07\\
         Seg I& 24 & 32.0 & 0 & 32.00\\
         Ant II & 2920 & 3893 & 0.38 & 3065.61\\
    \end{tabular}
    \caption{Radial data of all dSphs shown in this paper. The 2D projected half-light radius, $R_{\rm{e}}$, is taken from Table 1 in \citet{MNRAS:Simon2019}. We convert to 3D deprojected half-light radius, $r_{1/2}$ by multiplying by 4/3 \citep{Wolf:2009tu}. The ellipticity, $\epsilon$ is the Plummer model value from Table 3 in \citet{MNRAS:Munoz2018}. Finally, the sphericallized 3D $r_{1/2}$ is calculated as $\rm{3D }~r_{1/2} \sqrt{1 - \epsilon}$. Data for Ant II is taken from \citet{Torrealba:2019}. The first nine rows are the classical dSphs, and the last 10 rows are the ultra-faint dSphs that we use in this paper.}
    \label{tab:dsph}
\end{table}

\bsp	
\label{lastpage}
\end{document}